\newif\ifextended\extendedfalse
\newif\ifaec\aecfalse
\newif\ifreview\reviewfalse

\RequirePackage[svgnames,dvipsnames,prologue]{xcolor}
\PassOptionsToPackage{unicode}{hyperref}
\PassOptionsToPackage{naturalnames}{hyperref}

\ifreview
\documentclass[authoryear,acmsmall,screen,anonymous,review]{acmart}
\else
\documentclass[authoryear,acmsmall,screen]{acmart}
\fi
\usepackage{higher}

\newcommand\Supplemental{\ifreview, included in the anonymized supplementary material\else \citep{artifact}\fi\xspace}

\title{Generic Programming with Extensible Data Types}
\subtitle{\textit{Or,} Making Ad Hoc Extensible Data Types Less Ad Hoc}

\author{Alex Hubers}
\orcid{0000-0002-6237-3326}
\affiliation{
  \department{Department of Computer Science}
  \institution{The University of Iowa}
  \streetaddress{14 MacLean Hall}
  \city{Iowa City}
  \state{Iowa}
  \country{USA}}
\email{alexander-hubers@uiowa.edu}

\author{J. Garrett Morris}
\orcid{0000-0002-3992-1080}
\affiliation{
  \department{Department of Computer Science}
  \institution{The University of Iowa}
  \streetaddress{14 MacLean Hall}
  \city{Iowa City}
  \state{Iowa}
  \country{USA}}
\email{garrett-morris@uiowa.edu}

\newcommand\secfig[2]{Figure \ref{fig:#2}, \S\ref{sec:#1}}

\newcommand\EqTyDef{%
  \TyC{Eq} &: \TypeK \to \TypeK \\%
  \TyC{Eq} &= \lambda t. t \to t \to \Bool}
\newcommand\EqSigDef{%
  \ExC{eq_\Sigma} &: \forall z \co \RowK \TypeK. \, \Pi (\TyC{Eq} \, z) \to \TyC{Eq} \, (\Sigma z) \\%
  \ExC{eq_\Sigma} &= \lambda d \, v \, w. \, \Ana \, (\lambda l \, y. (\ExC{case} \, l \, (\lambda x. \, \ExC{sel} \, d \, l \, x \, y) \Branch \ExC{const} \, \ExC{False}) \, v) \, w}

\setcopyright{rightsretained}
\acmPrice{}
\acmDOI{10.1145/3607843}
\acmYear{2023}
\copyrightyear{2023}
\acmSubmissionID{icfp23main-p40-p}
\acmJournal{PACMPL}
\acmVolume{7}
\acmNumber{ICFP}
\acmArticle{201}
\acmMonth{8}
\received{2023-03-01}
\received[accepted]{2023-06-27}



\begin{document}

\begin{abstract} We present a novel approach to generic programming over
extensible data types.  Row types capture the structure of records and variants,
and can be used to express record and variant subtyping, record extension, and
modular composition of case branches.  We extend row typing to capture generic
programming over rows themselves, capturing patterns including lifting
operations to records and variations from their component types, and the duality
between cases blocks over variants and records of labeled functions, without
placing specific requirements on the fields or constructors present in the
records and variants.  We formalize our approach in System~\RO, an extension of
\FO with row types, and give a denotational semantics for (stratified) \RO in
Agda.
\end{abstract}

\maketitle

\section{Introduction}

The goal of extensible data types is to bring type safety to modular software development.  Row types~\citep{Wand87,Remy92} are one approach to that goal.  Rows express the structure of records or variants; row polymorphism captures properties like subtyping while maintaining a purely parametric approach to typing.  Row typing was originally designed to model object-oriented inheritance, but its applications include: extensible variants in OCaml~\citep{Garrigue98}; extensible effects~\citep{LindleyC12}; typing algebraic effects and handlers~\citep{HillerstromL16,Leijen14,Leijen17}; and, extensible protocols in session types~\citep{LindleyM17}.

This paper explores generic programming over rows.  Consider defining equality functions for extensible records.  Of course, given a particular set of fields, and knowledge of how to compare the field types, existing row type systems can express the equality function for records of those fields.  Even with metaprogramming support, however, having to explicitly define equality functions for each record type (and each extension of a record type) creates a significant burden for programmers---a disadvantage for an approach designed to encourage this style of programming!  Moreover, approaches that depend on particular sets of fields cannot extend to row polymorphism, a key contributor to the expressiveness of row types.  While we could express the extension of a particular record type, we could not (modularly) express that such an extension supports equality.

We propose novel record and variant operations, generic in the particular labels that appear in those records and variants, and realize these operations in System \RO, a core calculus that extends System \FO with row types based on \Rose~\citep{MorrisM19}.  Consider the equality function for extensible variants: if we know how to compare the values at each constructor in two variants, we ought to know how to compare the variants.  \Cref{fig:comparing-extensible-variants} captures this idea in \RO; we have elided type abstractions, applications, and annotations on bound variables, as they can be inferred from the given type signatures.  Type operator $Eq$ maps types to equality operators for those types.  Function $eq_\Sigma$ compares two variant values $v$ and $w$, given a record $d$ of comparison operators for their fields.  Suppose that $z$ is instantiated with the row $\Row{\LabTy {\Lab a} {\Int}, \LabTy {\Lab b} {\List \, \Bool}}$: $d$ will be a record of comparison functions $\Pi \Row{\LabTy {\Lab a} {\Int \to \Int \to \Bool}, \LabTy {\Lab b} {\List\,\Bool \to \List\,\Bool \to \Bool}}$, and $v$ and $w$ will each be variants $\Sigma \Row {\LabTy {\Lab a} {\Int}, \LabTy {\Lab b} {\List \, \Bool}}$.  (We follow \citet{PottierR05} in implicitly lifting operators on types, like $Eq$, to the corresponding operators on rows.)  Our key novelty is the $ana$ combinator: $\Ana \, f \, w$ \emph{ana}lyzes variant $w$, calling $f$ with its constructor label $l$ and contents $y$.  With these in hand, we can then rely on the variant branching combinator ($\Branch$) of \Rose: in $case$ $v$ is constructed with label $l$, we \emph{sel}ect from $d$ the $l$-labeled function, and use it to compare the contents $x$ and $y$ of the two variants; otherwise, regardless of the contents of $v$, we can return $\ExC {False}$.  We will return to each component of this definition in the remainder of the paper.

\begin{figure}
\small
\[
\begin{aligned}
\EqTyDef
\end{aligned}
\quad\vrule\quad
\begin{aligned}
\EqSigDef
\end{aligned}
\]
\caption{Comparing extensible variants in \RO}
\label{fig:comparing-extensible-variants}
\end{figure}

The generic operations in \RO build on the row type theory \Rose. \Rose is distinguished from other row type theories by two features.  First, \Rose uses qualified types~\citep{Jones94} to capture the structure of row types, rather than incorporating the structure of rows directly into the types of records and variants.  This indirection makes it possible to capture structural invariants in \Rose that are difficult or impossible to capture in other row type systems.  We rely on this expressiveness in typing the combinators in \RO: the function argument to $\Ana$, for example, is typed given the assumption that $l$ labels a value of type $u$ in row $z$.  Second, \Rose builds on a general account of rows as partial monoids, encompassing a variety of different row type theories in the literature.  While we will fix a particular theory of rows in our formalization of \RO, we will show how our account would generalize other theories of rows as well.

One way to realize the behavior of $eq_\Sigma$ would be to treat (hashes of) labels as keys at runtime: variants would be labeled by these keys and records would be dictionaries over keys.  While direct, this approach is neither practically nor theoretically sound: we would be disappointed to learn that selecting a field from a record was not a constant time operation, and it would be difficult to show that well-typed operations only relied on keys that were dynamically present in records or case blocks. We will show that \RO has a type-safe implementation with no runtime comparison or manipulation of labels, guided by the use of predicates in the types of the generic operations.  To do so, we will give a denotational semantics for universe-stratified \RO typing derivations in Agda, in which rows are interpreted as functions from finite naturals to types, row inclusions and combinations are witnessed by maps from finite naturals to finite naturals, and we have static guarantees that our indexing of records and variants is well-typed.

To summarize, this paper contributes:
\begin{itemize}
\item The extension of \Rose to generic programming over rows~\cref{sec:generic}, particularly the design of combinators that express generic transformations of row-typed products and sums;
\item A formalization of our approach in the \RO calculus~\cref{sec:ro}, which extends System~\FO with \Rose-style row typing, first-class labels, and generic programming over rows; and,
\item The denotation of \RO derivations in Agda~\cref{sec:semantics}, showing that \RO is sound and need not introduce runtime manipulation or comparison of labels.
\end{itemize}

We begin with a review of extensible datatypes in \Rose~\cref{sec:background} and conclude with discussions of related~\cref{sec:related} and future~\cref{sec:conclusion} work.

\section{Extensible Datatypes and the \Rose Type Theory}
\label{sec:background}

The goal of row typing is to support type-safe extensible data types.  This section gives an intuitive overview of row typing, and the \Rose type system in particular, preparatory to its extension in the following section.

\subsection{The Need for Extensibility}
\label{sec:motivation}

Existing functional language type theories are remarkably expressive, and further additions are rightly viewed with some suspicion.  We begin with two examples of the additional value of extensible data types.

\paragraph{The expression problem.}

\citet{Wadler98} describes the expression problem as ``a new name for an old problem''.  Consider an abstract data type along with several operations.  For example, we could have a simple type for arithmetic expressions, consisting of constants and sums, along with an operations to reduce expressions to integer values.  The challenge  is to extend this in two dimensions---say, by adding a new constructor for products, and a new operation to print expressions as character strings---without rewriting or recompiling existing code, and without compromising type safety.  In modern functional languages, adding new operations is easy, but adding new constructors requires changing the original type definition and all the existing definitions.  In object-oriented languages, adding new cases is easy but adding new operations requires changing the base class and all of its inheritors.  Programmers in either camp must resort to encoding tricks to capture the remaining case, making code more difficult to read and maintain.

\paragraph{Modular transformations.}

The expression problem may not seem entirely compelling: why artificially restrict a common refactoring operation to preserve existing code?  As an alternative view of the same problem, consider desugaring or optimization passes in a compiler.  We might hope to limit many of our passes to operating on a subset of the whole language~\citep{SarkarWD04,KeepD13}: for example, a single pass might resolve infix applications, while not changing the remainder of the syntax tree.  In writing these passes, we would like to make them generic over the untouched (or only recursively transformed) parts of the syntax tree.  This will both make the compiler more readable and maintainable, and provide type-based guarantees of the limited scope of these passes.  This problem is essentially the dual of the expression problem: instead of planning to extend our AST, we hope to write passes without fixing most of the AST.

\subsection{Row Types}

Consider the type of a function that selects the field $\Lab x$ from a record---we might write this function $\lambda r. sel \, r \, \Lab x$, where $sel$ is our function for \emph{sel}ecting record fields, and we use the teletype font to distinguish label constants from label variables.  We can imagine many record types which might contain an $\Lab x$ field---points on a plane, or in space; pixels on a screen; nodes for lambda expressions in the AST of a functional programming language---and in each case, $\Lab x$ might have a different type.  A general type for this function ought to encompass all its possible arguments, associating each with the corresponding result type.  This problem, along with its dual for variants, is the starting point for row type systems.

\subsubsection*{Rows and row polymorphism.}

A row is an association of labels to types.  For example, we write $\Row {\LabTy {\Lab x} Double, \LabTy {\Lab y} Double}$ for the row that associates both the labels $\Lab x$ and $\Lab y$ with the type of double-precision floating point numbers.  Record and variant types are constructed from rows; for example, a type for Cartesian coordinates is $\Pi \Row {\LabTy {\Lab x} Double, \LabTy {\Lab y} Double}$, while a more general type for points is $\Sigma {\Row {\LabTy {\Lab {Cart}} {\Pi {\Row {\LabTy {\Lab x} Double, \LabTy {\Lab y} Double}}}, \LabTy {\Lab {Polar}} {\Pi {\Row {\LabTy {\Lab r} Double, \LabTy {\Lab {theta}} Double}}}}}$.

Just introducing rows gets us little closer to solving our initial problem: we can say that $\lambda r. sel \, r \, \Lab{x}$ could have type $\Pi {\Row {\LabTy {\Lab x} Double}} \to Double$ or $\Pi {\Row {\LabTy {\Lab x} Int, \LabTy {\Lab y} Int}} \to Int$, but these are not a general account of its behavior.  Instead, this function should have a polymorphic type.  In many row type systems~\citep{Wand87,Remy89}, its type would be written similarly to $\forall t \, z. \,\Pi \Row {\LabTy {\Lab x} t \mid z} \to t$.  The syntax $\Row {\LabTy {\Lab x} t \mid z}$ denotes the \emph{extension} of row $z$ with the field $\LabTy {\Lab x} t$.  As a whole, the type denotes a function from a record containing any fields $z$, and also $\LabTy {\Lab x} t$, to a value of type $t$.  Similarly, a function that added a new $\Lab x$ field to an existing record could be given a type like $\forall t \, z. \, t \to \Pi z \to \Pi \Row{\LabTy {\Lab x} t \mid z}$.

This account of row types leaves several questions.  First: in the types above, can the instantiation of $z$ already include an association for $\Lab x$? 

\begin{itemize}[wide,nosep]
\item \citet{Wand87} allows free instantiation of $z$; extension is then interpreted as \emph{overwriting} the existing meaning of fields (in both types and terms).
\item \citet{Remy89} uses the kind system to preclude conflicting meaning of fields, but must introduce a new kind to capture functions which can either overwrite or extend objects.
\item \citet{BerthomieuM95} and \citet{Leijen05} allow free instantiation of $z$, and interpret extension as \emph{shadowing} the existing meaning of fields, such that the original meaning can be recovered later.
\end{itemize}

Second: does this account generalize from single field extension to arbitrary concatenation of objects?  For example, given two records, one of location data and one of color data, can we combine them to form a single record of colored location (or located color) data?

\subsubsection*{Polymorphism and predicates.}

\citet{Wand89} proposes the following term as a test of row type systems with record concatenation:
\[
  \lambda m \, n. \, \ExC {sel} \, (m \Concat n) \, {\Lab k}
\]
Here $m$ and $n$ are arbitrary records, and the function projects the field $\Lab k$ from their concatenation.  (In Wand's original example, $m$ and $n$ were records of method implementations, $\Lab k$ is a method name, and the term as a whole models multiple inheritance.)  The crux of the problem is that if we have to assign a type to either $m$ or $n$ that already commits to field $\Lab k$, then we have over-specified the behavior of the function.  On the other hand, if we do not commit to either $m$ or $n$ containing field $\Lab k$, then how can we be sure the function is well-defined at all?

This is the starting point for the \Rose type theory~\citep{MorrisM19}.  Instead of capturing the structure of rows directly in the types of records and variants, \Rose captures them using predicates in qualified types.  For example, in \Rose, the type of the $\Lab x$-selection function would be expressed as
\[
  \forall t \, z. \, \Leqp {\Row{\LabTy {\Lab x} t}} z \then \Pi z \to t
\]
That is: this is a function that maps $z$-shaped records to $t$ results, for any types $t$ and $z$, \emph{such that} the singleton row $\Row {\LabTy {\Lab x} t}$ is contained in $z$.  \Rose supports concatenation of records via predicates as well.  The type for Wand's example in \Rose is:
\[
  \forall t \, z_1 \, z_2 \, z_3. \, (\RowPlusP {z_1} {z_2} {z_3}, \Leqp {\Row {\LabTy {\Lab k} t}} {z_3}) \then \Pi z_1 \to \Pi z_2 \to t
\]
That is: this is a function that maps a $z_1$-shaped record and a $z_2$ shaped record to a $t$ result, \emph{such that} $z_1$ and $z_2$ can be concatenated to give row $z_3$, and $z_3$ contains the singleton row $\Row {\LabTy {\Lab k} t}$.  This type captures the full generality of Wand's challenge: we do not overconstrain either $m$ or $n$ to always provide field $\Lab k$, but still guarantee that the projection will always be well-defined.

\begin{figure}
\begin{small}
\begin{gather*}
\ib{\irule{\TypeJ \Gamma M \tau};
          {\TypeJ \Gamma {\LabTerm \ell M} {\LabTy \ell \tau}}}
\rsp
\ib{\irule{\TypeJ \Gamma M {\Pi {\rho_1}}}
          {\EntJ \Gamma {\Leqp {\rho_2} {\rho_1}}};
          {\TypeJ \Gamma {\Prj M} {\Pi {\rho_2}}}}
\rsp
\ib{\irule{\TypeJ \Gamma {M_1} {\Pi {\rho_1}}}
          {\TypeJ \Gamma {M_2} {\Pi {\rho_2}}}
          {\EntJ \Gamma {\RowPlusP {\rho_1} {\rho_2} {\rho_3}}};
          {\TypeJ \Gamma {M_1 \Concat M_2} {\Pi \rho_3}}}
\\
\ib{\irule{\TypeJ \Gamma M {\LabTy \ell \tau}};
          {\TypeJ \Gamma {\Unlabel M \ell} {\tau}}}
\rsp
\ib{\irule{\TypeJ \Gamma M {\Sigma {\rho_1}}}
          {\EntJ \Gamma {\Leqp {\rho_1} {\rho_2}}};
          {\TypeJ \Gamma {\Inj M} {\Sigma {\rho_2}}}}
\rsp
\ib{\irule{\TypeJ \Gamma {M_1} {\Sigma {\rho_1} \to \tau}}
          {\TypeJ \Gamma {M_2} {\Sigma {\rho_2} \to \tau}}
          {\EntJ \Gamma {\RowPlusP {\rho_1} {\rho_2} {\rho_3}}};
          {\TypeJ \Gamma {M_1 \Branch M_2} {\Sigma {\rho_3} \to \tau}}}
\end{gather*}
\end{small}
\caption{Typing of record and variant operations in \Rose}
\label{fig:rose-records-variants}
\end{figure}

\Cref{fig:rose-records-variants} gives the typing rules for the record and variant operations in \Rose.  The projection and injection operators are the generalizations of record selection and variant construction; each relies on being able to prove that one row is contained in another.  Branching ($M_1 \Branch M_2$) is dual to record concatenation ($M_1 \Concat M_2$): it combines eliminators for two variants to give the eliminator for their combination.  As with concatenation, it relies on being able to prove that the two smaller rows can be combined.

\subsubsection*{Theories of rows}

As \Rose does not commit directly to the structure of rows, but abstracts their structure via the \emph{containment} ($\Leqp {\rho_1} {\rho_2}$) and \emph{combination} ($\RowPlusP {\rho_1} {\rho_2} {\rho_3}$) predicates, it can be adapted to any of the different notions to row extension:

\begin{itemize}[wide,nosep]
\item To capture \emph{non-overlapping} rows: we stipulate that $\RowPlusP {\rho_1} {\rho_2} {\rho_3}$ is only satisfiable when $\rho_1$ and $\rho_2$ have no fields in common.  For this approach, we can define $\Leqp {\rho_1} {\rho_2}$ to hold either when there is some $\rho'$ such that $\RowPlusP {\rho_1} {\rho'} {\rho_2}$ or when $\RowPlusP {\rho'} {\rho_1} {\rho_2}$.

\item To capture \emph{overwriting}: $\RowPlusP {\rho_1} {\rho_2} {\rho_3}$ is always satisfiable, where $\rho_3$ reflects $\rho_1$ for any labels that appear in both.  We can define $\Leqp {\rho_1} {\rho_2}$ to hold exactly when there is $\rho'$ such that $\RowPlusP {\rho_1} {\rho'} {\rho_2}$.  (On the other side, when $\RowPlusP {\rho'} {\rho_1} {\rho_2}$, we cannot necessarily recover fields in $\rho_1$ from the combination $\rho_2$ because they may have been overwritten by fields in $\rho'$.)

\item To capture \emph{shadowing}: $\RowPlusP {\rho_1} {\rho_2} {\rho_3}$ is always satisfiable, and we get two containment predicates, $\Leqp [\Left] {\rho_1} {\rho_2} \iff \RowPlusP {\rho_1} {\rho'} {\rho_2}$ and $\Leqp [\Right] {\rho_1} {\rho_2} \iff \RowPlusP {\rho'} {\rho_1} {\rho_2}$, with corresponding injection and projection functions.
\end{itemize}

\Rose itself is defined generically over a \emph{row theory}, which defines the underlying structure of rows and interpretation of row predicates.  So,  \Rose encompasses all of the above cases, as well as both simpler (e.g., unlabeled) and more complex (e.g. modules) cases.

\subsection{Open Problems in Extensibility}

Despite \Rose's expressiveness, it is still limited in how it describes individual rows.  \Rose can capture the structure of rows, but it has no predicates that capture properties of the types in a row.  This limitation has several consequences.

If we know that every type in a variant or record supports equality comparisons, we should expect that the variant or record supports equality comparison as well.  However, even expressing this problem is not possible in \Rose---the constraint we need to express is on the types that appear in the row, not on the structure of the row itself.  The problem recurs when considering higher-order polymorphism.  Recall the example of modular AST transformations~\cref{sec:motivation}.  To maximize flexibility and readability, the pass that transforms infix to prefix applications should not constrain the remainder of the syntax tree.  However, this transformation is not only applied at the top level of expressions or definitions; it must also be applied recursively, regardless of the other nodes in the AST.  This, in turn, implies some constraint (such as functoriality) on the remainder of the AST, which cannot be captured in \Rose.

We know that records and variants enjoy strong duality properties: a case expression eliminating a variant corresponds to a record of functions, containing one (appropriately typed) function for each branch in the case expression.  This duality is not just of theoretical interest.  For example, in implementing a system of algebraic effects and handlers~\citep{PlotkinP03,PlotkinP09}, we could represent effectful computations as abstract syntax trees over operations, and handlers as records of implementations of those operations.  We might then hope to define a general handling combinator, which combines an effectful computation with an appropriate handler.  However, we cannot implement this operation in \Rose: while we can use the same row variable to describe both records and variants (i.e., both computations and their handlers), the branching and projection operators all refer to specific labels.

Existing row type theories address some of these problems.  \citet{BlumeAC06} distinguishes case blocks from functions, and realizes case blocks by records of functions in their semantics.  However, this step in the semantics is not available to programmers.  \citet{PottierR05} implicitly lift operations on types to operations on rows: if $z$ is a row of associations $\LabTy {\ell_i} {\tau_i}$, then $z \to \upsilon$ is the row of associations $\LabTy {\ell_i} {\tau_i \to \upsilon}$.  They further postulate an operation $rapply$ which applies a record of functions to a record of (identically labeled) arguments, producing a record of results.  However, this operation is treated as a primitive extension of their calculus.  \citet{Chlipala10} includes a mapping operator on records in a calculus based on \FO, generalizing the lifting of Pottier and \Remy, and provides type-directed generation of record folding operations.  He does not consider variants in his approach; moreover, it is not immediately clear that folds, and their duals for variants, would be sufficient to capture the open problems we identify.

\section{Generic Programming in \RO}
\label{sec:generic}

System \RO generalizes \Rose in two dimensions.  \Rose imposes Hindley-Milner constraints on typing; \RO is based on System~\FO extended with qualified types, and so supports first-class polymorphism and general type operators.  More significantly, the record and variant operations in \Rose are all specific to concrete labels or sets of labels; \RO introduces label-generic combinators.  This section introduces \RO by example.

Through the majority of this section, we will assume \emph{simple rows}: labels are restricted to appear at most once in a given row, row combination is commutative (and so there is a single containment operator), and $\RowPlusP {\rho_1} {\rho_2} {\rho_3}$ is unsatisfiable if $\rho_1$ and $\rho_2$ contain any of the same fields.  This is the most common approach to typing records and variants and rules out many unexpected behaviors.  At the end of the section~\cref{sec:non-commutative}, we will discuss the specific challenges in extending our development to a non-commutative row theory.

\subsection{First-Class Labels}
\label{sec:first-class-labels}




In \Rose, labels exist in types, but not in terms.  The construction ($\LabTerm \ell M$) and destruction ($\Unlabel M \ell$) terms, which are overloaded for both singleton records and variants, are each essentially infinite families of terms, one for each label.  To support label-generic operations, however, we will need to make labels first-class citizens in the term language as well as the type language.

To do so, we follow the approach used by~\citet{GasterJ96} and \citet{Sulzmann97}.  We have added a singleton type constructor $\Sing -$ to \RO: if $\ell$ is a label type, then $\Sing \ell$ is the corresponding singleton type.  (For a label constant $\Lab{L}$, we also write $\Lab L$ for the unique inhabitant of $\Sing {\Lab L}$.)  First-class labels allow us to abstract several common patterns in \Rose.  For example, to select an individual field from a record, we first apply $\Prj$ to project a singleton record and then use the singleton deconstruction operator.  \Rose introduced syntactic sugar for this pattern; in contrast, we can define the selection function directly in \RO by:
\begin{align*}
  \ExC{sel} &: \forall l \co \LabK, t \co \TypeK, z \co \RowK \TypeK. \, \Leqp {\Row {\LabTy l t}} z \then \Pi z \to \Sing l \to t \\
  \ExC{sel} &= \Lambda (l \co \LabK) \, (t \co \TypeK) \, (z \co \RowK \TypeK). \, \lambda (r \co \Pi z) \, (g : \Sing l). \, \Unlabel {\Prj r} g
\intertext{(Note that predicate abstraction and application remain implicit in \RO.)  The type abstractions and annotations in this example, and most of the following, can be determined from the type signatures alone, so we will generally omit them:}
  \ExC{sel} &= \lambda r \, l. \, \Unlabel {\Prj r} l
\end{align*}

Row type systems are frequently forced to distinguish between record extension (which adds new fields to existing records) and record update (which changes the value---and possibly type---of an existing field in a record), because their types impose different requirements on the input record type.  \citet{Remy89} introduces \emph{presence polymorphism}, allowing a single term to play both roles at the cost of additional type system complexity.  A single term that captures both in \RO:
\begin{align*}
  \ExC{upd} &: \forall l \co \LabK, t, u \co \TypeK, z_1, z_2 : \RowK \TypeK. \, \Leqp {z_1} {\Row {\LabTy l t}} \then \Sing l \to u \to \Pi(\RowPlus {z_1} {z_2}) \to \Pi(\RowPlus {\Row {\LabTy l u}} {z_2}) \\
  \ExC{upd} &= \lambda l \, u \, r. \, (\LabTerm l u) \Concat {\Prj r}
\end{align*}
We treat $\odot$ as a \emph{partial} type constructor~\citep{JonesD08,IngleHM22}: we write $\RowPlus {\rho_1} {\rho_2}$ as a type to denote a fresh type variable $z$ under the constraint $\RowPlusP {\rho_1} {\rho_2} z$.  Row $z_1$ is either the empty row or the singleton row mapping $l$ to $t$; row $z_2$ is constrained to combine with $\Row{\LabTy l t}$, so cannot contain label $l$.  The input record, of type $\Pi(\RowPlus {z_1} {z_2})$ may contain field $l$ (depending on the choice of $z_1$); the output record definitely contains $l$, mapped to type $u$.  

\newcommand\BoolRow[1]{\Row{\LabTy{\Lab{True}}{#1}, \LabTy{\Lab{False}}{#1}}}
\newcommand\UnitTy{\Pi \Row{}}

First-class labels are also useful for capturing programming patterns with variants.  We can define a generic function for constructing variants:
\begin{align*}
  \ExC{con} &: \forall l \co \LabK, t \co \TypeK, z \co \RowK \TypeK . \, \Leqp {\Row {\LabTy l t}} z \then \Sing l \to t \to \Sigma z \\
  \ExC{con} &= \lambda l \, x. \, \Inj {(\LabTerm l x)}
\end{align*}
The base case for the branching operator $\Branch$ is a function that maps a singleton variant to a result.  We can capture this pattern as well:
\begin{align*}
  \ExC{case} &: \forall l \co \LabK, t \co \TypeK, u \co \TypeK. \, \Sing l \to (t \to u) \to \Sigma {\Row {\LabTy l t}} \to u \\
  \ExC{case} &= \lambda l \, f \, x. \, f \, (\Unlabel x l)
\end{align*}
Representing Booleans as $\TyC{Bool} = \Sigma \BoolRow {\UnitTy}$ (syntactic sugar for $\Sigma (\RowPlus {\Row{\LabTy {\Lab{True}} {\Pi \Row{}}}} {\Row{\LabTy {\Lab{False}} {\Pi \Row{}}}})$), we could then define the usual conditional by:
\begin{align*}
  \ExC{ifte} &: \forall t \co \TypeK. \, \TyC {Bool} \to t \to t \to t \\
  \ExC{ifte} &= \lambda b \, t \, f. \, (\ExC {case} \, {\Lab {True}} \, (\lambda u. \, t) \Branch \ExC {case} \, {\Lab {False}} \, (\lambda u. \, f)) \, b
\end{align*}

Perhaps most surprisingly, while \Rose lacked syntax or types for first-class labels, adding them does not require extending its semantics in any non-trivial way.  The necessary information for the \emph{sel} function, for example, is already captured entirely by the predicate $\Leqp {\Row {\LabTy l t}} z$.  The value of type $\Sing l$ provides no additional information---as you would expect for a value of a singleton type!

\subsection{The Duality of Records and Variants}
\label{sec:reify-reflect}

We begin our exploration of generic programming over rows with the duality between records and variants.  This duality is foundational to row type systems in general, and to \Rose in particular.  Its introduction rule for variants and the elimination rule for records are clearly dual, and the rules for concatenating variant eliminators and concatenating records are nearly as evidently dual.  (To make the duality more explicit, one could have defined a rule for combining record constructors---from $\tau \to \Pi z_1$ and $\tau \to \Pi z_2$, obtain $\tau \to \Pi (\RowPlus {z_1} {z_2})$---but this seems to obtain theoretical elegance at the cost of usability.)  In fact, we can witness this duality in \Rose, but only for concrete rows.  For example, we can define the following operations for the Boolean type:
\begin{smalle}
\begin{align*}
  \TyC{Cases_{\TyC{B}}} &: \TypeK \to \RowK \TypeK \\
  \TyC{Cases_{\TyC{B}}} &= \lambda t. \BoolRow{\UnitTy \to t} \\[1ex]
  \ExC{reify_{\TyC{B}}} &: \forall t \co \TypeK. \, (\TyC{Bool} \to t) \to \Pi (\TyC{Cases_{\TyC{B}}} \, t) \\
  \ExC{reify_{\TyC{B}}} &= \lambda f. \, (\LabTerm {\Lab{True}} {f \, (\ExC{con} \, \Lab{True} \, ())}) \Concat (\LabTerm {\Lab{False}} {f \, (\ExC{con} \, \Lab{False} \, ())}) \\[1ex]
  \ExC{reflect_{\TyC{B}}} &: \forall t \co \TypeK. \, \Pi (\TyC{Cases_{\TyC{B}}} \, t) \to \TyC{Bool} \to t \\
  \ExC{reflect_{\TyC{B}}} &= \lambda d. \, (\ExC{case} \, \Lab{True} \, (\ExC{sel} \, r \, \Lab{True})) \Branch (\ExC{case} \, \Lab{False} \, (\ExC{sel} \, r \, \Lab{False}))
\end{align*}
\end{smalle}
The type $Cases_B$ abbreviates operations over the constructors of the Boolean type.  The $reify_B$ function transforms a function that scrutinizes a Boolean value into a record of functions, one for the $\Lab{True}$ case and one for the $\Lab{False}$ case; dually, the $reflect_B$ function uses such a record of functions to scrutinize a Boolean value.  (We write $()$ for the unique value of the $\Pi\Row{}$ type.)  Knowing the constructors of the Boolean type is essential to writing this example; while such functions exist for any variant type in \Rose, their definition would have to be repeated for each type.

\begin{figure}
\small

\[
\begin{aligned}
  \ExC{reify} &: \forall z \co \RowK \TypeK, t \co \TypeK. \, (\Sigma z \to t) \to \Pi (z \to t) \\
  \ExC{reify} &= \lambda f. \, \Syn \, (\lambda l \, x. \, f \, (\ExC{con} \, l \, x))
\end{aligned}
\quad\vrule\quad
\begin{aligned}
  \ExC{reflect} &: \forall z \co \RowK \TypeK, t \co \TypeK. \, \Pi (z \to t) \to (\Sigma z \to t) \\
  \ExC{reflect} &= \lambda d \, w. \, \Ana \, (\lambda l \, u. \, \ExC{sel} \, d \, l \, u) \, w
\end{aligned}
\]

\caption{Witnessing the duality of records and variants}
\label{fig:reify-reflect}
\end{figure}
  
In \RO, we can write generic versions of these operators, applicable to any variant type and the corresponding record of cases, as shown in \Cref{fig:reify-reflect}.  The types of $reify$ and $reflect$ rely on lifting operations on types to operations on rows: if $z$ is the row of types $\LabTy {\ell_i} {\tau_i}$, then $z \to t$ is the row of types $\LabTy {\ell_i} {\upsilon_i \to t}$. In $reify_B$ and $reflect_B$, we relied on concrete constructors in two places: when deconstructing a Boolean value in $reflect_B$, and when building the record of constructors in $reify_B$.  \RO provides label-generic versions of these two operations, one for \emph{ana}lyzing variants and a dual operator for \emph{syn}thesizing records.  Here is our first attempt at their typing rules:
\begin{smalle}
\begin{gather*}
  \ib{\irule[\trule{${\Ana}_1$}]
  {\KindJ \Gamma \rho {\RowK \TypeK}}
  {\TypeJ \Gamma M {\forall l \co \LabK, u \co \TypeK. \Leqp {\Row {\LabTy l u}} \rho \then \Sing l \to u \to \tau}};
  {\TypeJ \Gamma {\Ana \, M} {\Sigma \rho \to \tau}}}
\\
\ib{\irule[\trule{${\Syn}_1$}]
  {\KindJ \Gamma \rho {\RowK \TypeK}}
  {\TypeJ \Gamma M {\forall l \co \LabK, u \co \TypeK. \Leqp {\Row {\LabTy l u}} \rho \then \Sing l \to u}};
  {\TypeJ \Gamma {\Syn \, M} {\Pi \rho}}}
\end{gather*}
\end{smalle}
We write $\RowK \kappa$ for the kind of rows over types of kind $\kappa$.  To avoid a sea of metavariables, we combine kinding and typing assertions in $\Gamma$; the judgment $\KindJ \Gamma \rho {\RowK \TypeK}$ is a kinding assertion on $\rho$, and $\TypeJ \Gamma {\Ana \, M} {\Sigma \rho \to \tau}$ is a typing assertion on $ana \, M$.

In $ana \, M$, the body $M$ is a label-generic version of the cases in a branch expression: given a label $l$, a type $u$, and evidence that $\Row{\LabTy l u}$ appears in $\rho$, $M$ consumes a single case---(a witness for) the constructor, and its contents---and produces a result of type $\tau$.  If $M$ can do so for any constructor appearing in $\rho$, then $ana \, M$ can consume a value of $\Sigma \rho$ to produce a result of type $\tau$.  We use $ana$ in implementing $reflect$.  Given the constructor label $l$ and contents $u$ of an arbitrary variant value $w$, we invoke the $l$-labeled entry from the record $d$ with argument $u$.  Again, lifting plays a central role: from $\Leqp {\Row {\LabTy l u}} z$, we can conclude that $\Leqp {\Row{\LabTy l {u \to t}}} {z \to t}$, and so $sel \, d \, l$ is a $u \to t$ function.

In $\Syn \, M$, the body $M$ is a label-generic version of the components of a concatenation expression: given a label $l$, a type $u$, and evidence that $\Row{\LabTy l u}$ appears in $\rho$, $M$ produces a value of type $u$.  If $M$ can do so for each label appearing in $\rho$, then $syn \, M$ can produce a record of type $\Pi \rho$.  We use $syn$ in implementing $reify$.  In the body, we have access to $f : \Sigma z \to t$.  We build a new function $u \to t$, which wraps its argument in constructor $l$ and then invokes $f$.  Lifting plays a similar role to its role in $reflect$: as $\Leqp {\Row {\LabTy l u}} z$, the result type includes $\LabTy l {u \to t}$.

\subsection{Transformations}
\label{sec:maps}

Next, we consider generic transformations on extensible types.

\subsubsection*{Type-preserving maps.}

We begin with type-preserving mappings, such as reversing each field of a record of lists.  Here is the version for records; the version for variants is nearly identical.
\begin{smalle}
\begin{align*}
  \MapP' &: \forall z \co \RowK \TypeK. (\forall l \co \LabK, u \co \TypeK. \, \Leqp {\Row {\LabTy l u}} z \then \Sing l \to u \to u) \to \Pi z \to \Pi z \\
  \MapP' &= \lambda f \, r. \, \Syn \, (\lambda l. \, f \, l \, (\ExC {sel} \, r \, l))
\end{align*}
\end{smalle}
The mapped function is label-generic: for any label $l$ and type $u$ appearing in $z$, the function transforms the old $u$ value into a new $u$ value.  Given such a function $f$ and a record $r$, we synthesize a new record in which each field $l$ contains the result of $f$ applied to the old field and its label. 

\subsubsection*{Type-transforming maps.}

The far more interesting case is type-transforming mappings, such as transforming a record of lists into a record of their lengths.  The challenge here is not defining the term (in fact, it will turn out to appear identical to the previous term), but rather to find an appropriately expressive type.  Suppose that we have type constructors $List : \TypeK \to \TypeK$ and $Int : \TypeK$, such that the $length$ function has type $\forall t \co \TypeK. \, List \, t \to Int$.  We might then imagine that the pointwise length function on records would have a type like
\[
  \forall z : \RowK \TypeK. \, \Pi (\List \, z) \to \Pi (\TyC {const} \, \Int \, z)
\]
where $const : \TypeK \to \TypeK \to \TypeK$ is the expected constant operator, on types.  In the input type, we lift the type constructor $List$ over the row $z$; this allows us to capture the idea of a row of list types.  In the output type, we lift $const \, Int$ over $z$; this replaces each type in $z$ by $Int$.  Instantiating this type with the concrete row $\Row {\LabTy {\Lab a} \Bool, \LabTy {\Lab b} \TyC{Char}}$ would give
\[
  \Pi \Row{\LabTy {\Lab a} {\List \, \Bool}, \LabTy {\Lab b} {\List \, \TyC{Char}}} \to \Pi \Row{\LabTy {\Lab a} \Int, \LabTy {\Lab b} \Int}
\]
Of course, we cannot inhabit this type with a term based on $\MapP'$, as the input and output types are not identical.  More seriously, however, it is not clear how we could inhabit it with any term based on our previous typing rule for $\Syn$.  The only types in the output row are $\Int$, and it is not clear how we could reconstruct an application of $length$ to a field of the input row given only the information that $\LabTy l \Int$ appears in the output row.

Our solution is to generalize the types of $\Ana$ and $\Syn$ to incorporate a type operator $\phi$:
\begin{smalle}
\begin{gather*}
\ib{\irule[\trule{${\Ana}_2$}]
          {\KindJ \Gamma \rho {\RowK {\shade\kappa}}}
          {\shade{\KindJ \Gamma \phi {\kappa \to \TypeK}}}
          {\TypeJ \Gamma M {\forall l \co \LabK, u \co \shade\kappa. \, \Leqp {\Row{\LabTy l u}} \rho \then \Sing l \to \shade{\phi \, u} \to \tau}};
          {\TypeJ \Gamma {\Ana [\shade\phi] \, M} {\Sigma \shade{(\phi \, \rho)} \to \tau}}}
\\
\ib{\irule[\trule{${\Syn}_2$}]
          {\KindJ \Gamma \rho {\RowK {\shade\kappa}}}
          {\shade{\KindJ \Gamma \phi {\kappa \to \TypeK}}}
          {\TypeJ \Gamma M {\forall l \co \LabK, u \co \shade\kappa. \, \Leqp {\Row{\LabTy l u}} \rho \then \Sing l \to \shade{\phi \, u}}};
          {\TypeJ \Gamma {\Syn [\shade\phi] \, M} {\Pi \shade{(\phi \, \rho)}}}}
\end{gather*}
\end{smalle}
Differences from the previous rules are shaded.  We now allow $\rho$ to range over rows of arbitrary kind $\kappa$---we will make use of this in capturing functoriality later in the section---and require that $\phi$ be a type operator mapping from $\kappa$ to $\TypeK$.  We then uniformly introduce $\phi$ in the uses of $\rho$, both in typing results of $\Ana$ and $\Syn$ and in typing their body.  Rules \trule{${\Ana}_1$} and \trule{${\Syn}_1$} are special cases of these rules, and going forward we will write $\Ana$ and $\Syn$ for $\Ana [\lambda t. \, t]$ and $\Syn [\lambda t. \, t]$, respectively.

\begin{figure}
\small  
\begin{align*}
  \KFam\Iter\kappa &: (\kappa \to \TypeK) \to (\kappa \to \TypeK) \to \RowK \kappa \to \TypeK \\
  \KFam\Iter\kappa &= \lambda f \, g \, z. \, \forall l \co \LabK, u \co \kappa. \, (\Leqp {\Row {\LabTy l u}} z) \then \Sing l \to f \, u \to g \, u \\[1ex]
  \KFam\MapP\kappa &: \forall z \co \RowK \kappa, f \co \kappa \to \TypeK, g \co \kappa \to \TypeK. \, \Iter \, f \, g \, z \to \Pi (f \, z) \to \Pi (g \, z) \\
  \KFam\MapP\kappa &= \Lambda z \, f \, g. \, \lambda i \, r. \, \Syn [g] \, (\lambda l. \, i \, l \, (\ExC{sel} \, r \, l)) \\[1ex]
  \KFam\MapS\kappa &: \forall z \co \RowK \kappa, f \co \kappa \to \TypeK, g \co \kappa \to \TypeK. \, \Iter \, f \, g \, z \to \Sigma (f \, z) \to \Sigma (g \, z) \\
  \KFam\MapS\kappa &= \Lambda z \, f \, g. \, \lambda i \, v. \, \Ana [f] \, (\lambda l \, x. \, \ExC{con} \, l \, (i \, l \, x)) \, v
\end{align*}

\caption{Transforming records and variants}
\label{fig:maps}
\end{figure}

With the generalized typing rules for $\Ana$ and $\Syn$, we can now define kind-indexed families type-transforming maps for record and variants, shown in \cref{fig:maps}.  We write $\KFam{\mathrm X}\kappa$ for a family of X's indexed by kind $\kappa$.  We would expect languages based on \RO to also include kind-polymorphism; we have omitted it from our formalization simply to avoid an orthogonal source of complexity.  The type $\KFam\Iter\kappa \, f \, g \, z$ captures iterated functions over row $z$; type operator $f$ is used to construct the input type, and $g$ is used to construct the output type.  We make the type abstractions in $\KFam\MapP\kappa$ and $\KFam\MapS\kappa$ explicit, as we will need to refer to the abstracted types in the calls to $\Syn$ and $\Ana$.  

The implementation of $\KFam\MapP\kappa$ is almost identical to the implementation of $\MapP'$.  The crucial difference is in providing the operator $g$ to $\Syn$.  This means that the body of $\Syn$ has the type
\[
  \forall l \co \LabK, u \co \kappa. \, \Leqp {\Row{\LabTy l u}} z \then \Sing l \to g \, u
\]
That is to say: knowing that $\LabTy l u$ appears in $z$, we must produce a value of type $g \, u$.  The assumption is sufficient to conclude that $\LabTy l {f \, u}$ appears in $f \, z$, and so $sel \, r \, l$ is a suitable input to the iterated function $i : \KFam{Iter}\kappa \, f \, g \, z$.

The implementation of $\KFam\MapS\kappa$ is the expected dual of the implementation of $\KFam\MapP\kappa$.  We annotate $\Ana$ with the input-side operator $f$, so its body has the type
\[
  \forall l \co \LabK, u \co \kappa. \, \Leqp {\Row{\LabTy l u}} z \then \Sing l \to f \, u \to \Sigma \, (g \, z)
\]
Here we are immediately sure that the value $x$ is a suitable input for $i$; from $\Leqp {\Row{\LabTy l u}} z$ we have $\Leqp {\Row{\LabTy l {g \, u}}} {g \, z}$, and so $con \, l \, (i \, l \, x)$ can be of type $\Sigma (g \, x)$.

\subsubsection*{Pointwise application.}

\citet{PottierR05} describe a pointwise-application operator for records, which maps a record of functions and record of arguments to a record of results.  We can describe a similar family of operators in \RO, as follows:
\begin{smalle}
\begin{align*}
  \KFam{\TyC{Xf}}\kappa &: (\kappa \to \TypeK) \to (\kappa \to \TypeK) \to (\kappa \to \TypeK) \\
  \KFam{\TyC{Xf}}\kappa &= \lambda f \, g \, a. \, f \, a \to g \, a \\[1ex]
  \KFam{\ExC{rapply}}\kappa &: \forall f \co \kappa \to \TypeK, g \co \kappa \to \TypeK, z \co \RowK \kappa. \Pi (\KFam{\TyC{Xf}}\kappa \, f \, g \, z) \to \Pi (f \, z) \to \Pi (g \, z) \\
  \KFam{\ExC{rapply}}\kappa &= \lambda d \, r. \, \KFam\MapP\kappa \, (\lambda l \, x. \, \ExC {sel} \, d \, l \, x) \, r
\end{align*}
\end{smalle}
The type $\KFam{\TyC{Xf}}\kappa$ describes the individual transformation functions; as we expect to have a record of these functions, suited to their record of arguments, we do not have to describe them in a label-generic way.  The $\KFam{\ExC{rapply}}\kappa$ function then takes a record of such transformers (note that we lift $\KFam{\TyC{Xf}}\kappa \, f \, g$ from an operator on $\kappa$ to an operator on $\RowK \kappa$) and a record of arguments, and produces a record of results.  Its implementation is a direct application of $\MapP$, in which the body need only look up the appropriately labeled function in the input $d$.

Our $rapply$ is not quite the same as Pottier and \Remy's: where we rely on type applications $f \, z$ and $g \, z$ based on a single row, they define a pointwise lifting of the function constructor to rows $z_1 \to z_2$.  However: their rows are infinite, with a default type for all labels not mentioned in the row; correspondingly, their records are infinite, with a default value for all labels not mentioned in building the record.  This means that $z_1 \to z_2$ can always be well-defined, by using $z_1$'s default type as the domain for any labels not mentioned in $z_1$ and $z_2$'s default type as the codomain for any labels not mentioned in $z_2$.  With finite rows, we do not have the same luxury.  Should we interpret $z_1 \to z_2$ as undefined if the label sets of $z_1$ and $z_2$ are not identical?  Or restrict it to the intersection of those label sets?  The former would introduce additional partiality in the type of $rapply$, while the latter would seem to make $rapply$ impossible to define.  Without a more compelling application of this additional flexibility, we have limited ourselves to lifting type operators over rows.

\subsubsection*{Lifting functoriality.}
\newcommand\FmapS{\ExC{fmap_\Sigma}}
\newcommand\FmapP{\ExC{fmap_\Pi}}

A more substantial application of the map functions is in lifting functoriality---as realized in languages like Haskell---to records and variants.  The idea is that if we have a row \emph{of type constructors}, where each constructor in the row has a suitable mapping operator, then we can derive mapping operators for record and variant type constructors built from that row.  Our implementation is shown in \cref{fig:functors}.

\begin{figure}
\small
\begin{gather*}
\begin{aligned}
  \TyC{Functor} &: (\TypeK \to \TypeK) \to \TypeK \\
  \TyC{Functor} &= \lambda f. \, \forall t \co \TypeK, u \co \TypeK. \, (t \to u) \to f \, t \to f \, u \\[1ex]
\end{aligned}
\\
\begin{aligned}
  \FmapS &: \forall z \co \RowK {\TypeK \to \TypeK}. \, \Pi (\TyC{Functor} \, z) \to \TyC{Functor} \, (\Sigma \, z) \\
  \FmapS &= \lambda d \, f \, w. \, \KFam\MapS{\TypeK\to\TypeK} \, (\lambda l \, x. \, \ExC {sel} \, d \, l \, f \, x) \, w\\[1ex]
\end{aligned}
\quad\vrule\quad
\begin{aligned}
  \FmapP &: \forall z \co \RowK {\TypeK \to \TypeK}. \, \Pi (\TyC{Functor} \, z) \to \TyC{Functor} \, (\Pi \, z) \\
  \FmapP &= \lambda d \, f \, r. \, \KFam\MapP{\TypeK\to\TypeK} \, (\lambda l \, x. \, \ExC {sel} \, d \, l \, f \, x) \, r
\end{aligned}
\end{gather*}

\caption{Lifting functoriality to records and variants}
\label{fig:functors}
\end{figure}

We begin by defining the $Functor$ type operator.  This should be read as capturing the evidence that a type operator is a functor: $Functor \, List$, for example, is $\forall t \, u. (t \to u) \to List \, t \to List \, u$. 

We turn to the types of $\FmapS$ and $\FmapP$.  We abstract over a row $z$ of type constructors.  Lifting $Functor$ over $z$ gives a row of types, so $\Pi (Functor \, z)$ is a record of evidence that each constructor in $z$ is functorial.  Now, we want to make a claim about record and variant types built from $z$.  To do so, we generalize $\Pi$ and $\Sigma$ to families of type constructors, where for $z : \kappa_1 \to \kappa_2$  we write $\Sigma z$ for the type constructor $\lambda t. \, \Sigma (z \, t)$ and similarly for $\Pi$.  This generalization is not necessary---we could write the constructors out---but this abbreviation seems intuitive, and makes the types of $\FmapS$ and $\FmapP$ natural.

Finally, we can implement $\FmapS$ and $\FmapP$ directly using $\MapS$ and $\MapP$; in each case, the mapped function simply looks up the appropriate evidence in $d$, then applies it to lift $f$ over $x$.

\subsection{Comparing Records and Variants}
\label{sec:comparisons}

We continue exploring component-wise operations on variants.  Our goal is to compare values of two variant types, given that we can compare the values at each of their constructors.  Our intended code in shown in \cref{fig:equality-variants}.  We begin by defining the type operator $Eq$, which captures equality comparisons (actually, any binary comparison); given a row $z$, $\Pi (Eq \, z)$ is a record of comparison operators for each type in $z$.  To compare two values $v, w$ of type $\Sigma z$, we begin by analyzing $w$.  We can then fall back on the branching combinator of \Rose: if $v$ is also built with constructor $l$, we can compare their contents using the $l$ field of $d$; otherwise, the two are definitely unequal.

\begin{figure}
\small
\[
  \begin{aligned}
  \EqTyDef
  \end{aligned}
  \quad \vrule \quad
  \begin{aligned}
  \EqSigDef 
  \end{aligned}
\]

\caption{Comparing values of variant type}
\label{fig:equality-variants}
\end{figure}

The only difficulty with this implementation is that it does not type.  Consider the branch expression in the body of $ana$.  As $v : \Sigma z$, we must show that the two branches combine to give $z$.  However, all we know is that $\Leqp {\Row {\LabTy l u}} z$; while logically this implies that there must be a ``remainder'' of $z$ less $\Row {\LabTy l u}$, we do not have access to it.

Our solution is to update the typing rules for $\Ana$ and $\Syn$, generalizing the type of the body.
\begin{smalle}
\begin{gather*}
\ib{\irule[\trule{${\Ana}_3$}]
          {\KindJ \Gamma \rho {\RowK \kappa}}
          {\KindJ \Gamma \phi {\kappa \to \TypeK}}
          {\TypeJ \Gamma M {\forall l \co \LabK, u \co \kappa, \shade{y \co \RowK \kappa}. \, \shade{\RowPlusP {\Row {\LabTy l u}} y \rho} \then \Sing l \to \phi \, u \to \tau}};
          {\TypeJ \Gamma {\Ana [\phi] \, M} {\Sigma (\phi \, \rho) \to \tau}}}
\\
\ib{\irule[\trule{${\Syn}_3$}]
          {\KindJ \Gamma \rho {\RowK \kappa}}
          {\KindJ \Gamma \phi {\kappa \to \TypeK}}
          {\TypeJ \Gamma M {\forall l \co \LabK, u \co \kappa, \shade{y \co \RowK \kappa}. \, \shade{\RowPlusP {\Row {\LabTy l u}} y \rho} \then \Sing l \to \phi \, u}};
          {\TypeJ \Gamma {\Syn [\phi] \, M} {\Pi (\phi \, \rho)}}}
\end{gather*}
\end{smalle}
The changed components of the rules are shaded.  Instead of providing evidence that $\Leqp {\Row {\LabTy l u}} \rho$, we now decompose $\rho$ into $\Row{\LabTy l u}$ and a row type $y$.  The previous iteration of the rule is a special case of this one.  With this rule, our intended implementation of $eq_\Sigma$ is well-typed.

Unfortunately, the solution for variants does not obviously dualize to give a comparison operator for records.  Again, assume we have comparators for each field.  The operators we have discussed so far would allow us to build a record of Booleans.  However, for the records to be equal, we must then determine whether those Booleans are all true, and (without knowing the specific fields) we have no tools to do so.

To capture functions like these, we introduce a folding operation over records:
\begin{smalle}
\[
\ib{\irule[\trule{$\FoldP$}]
          {\begin{array}{@{}c@{}}
             {M_1 : \forall l \co \LabK, t \co \TypeK, y \co \RowK {\TypeK}.  (\RowPlusP {\Row {\LabTy l t}} y \rho) \then \Sing l \to t \to \upsilon}
             \\
             {\TypeJ \Gamma {M_2} {\upsilon \to \upsilon \to \upsilon}}
             \isp
             {\TypeJ \Gamma {M_3} \upsilon}
             \isp
             {\TypeJ \Gamma N {\Pi \rho}}
           \end{array}};
          {\TypeJ \Gamma {\FoldP \, M_1 \, M_2 \, M_3 \, N} \upsilon}}
\]
\end{smalle}
The term $M_1$ is a label-generic mapping from the fields of the input record $N : \Pi \rho$ to the result type $\upsilon$; $M_2$ combines values of type $\upsilon$, and $M_3$ is an identity for $M_2$, used for folding the empty record.  Given this folding operator, we can define equality comparison for records, as shown in \cref{fig:equality-records}.

\begin{figure}
\small  
\begin{align*}
  \ExC{eq_\Pi} &: \forall z \co \RowK \TypeK. \, \Pi (\TyC{Eq} \, z) \to \TyC{Eq} \, (\Pi z) \\
  \ExC{eq_\Pi} &= \lambda d \, r. \, \FoldP \, (\lambda l \, x. \, (\ExC{sel} \, d \, l) \, (\ExC{sel} \, r \, l) \, x) \, (\&\&) \, \ExC{True}
\end{align*}  

\caption{Comparing values of record type}
\label{fig:equality-records}
\end{figure}

Introducing this operator immediately raises several questions.  For example: in what order are the mapped fields passed to the folding function?  Is the identity included once?  At all?  And so forth.  Our conclusion is that the values passed to $\FoldP$ must follow the same rules as the underlying row theory.  Following \citet{MorrisM19}, row theories must be associative and have the empty row as their unit; thus, $M_2$ should be associative and have $M_3$ as its unit.  For a commutative row theory (as we have been assuming), $M_2$ should be commutative as well.  For a non-commutative theory, on the other hand, $\FoldP$ would pass values to $M_2$ consistent with the ordering of fields in the row.  And so forth.  Absent these constraints, the exact behavior of $\FoldP$ ought to be unspecified.

The dual operator for variants would be an unfold, generating a variant by unrolling a starting value.  Introducing such an operator would raise all the same problems as we have for $\FoldP$.  As we have found no compelling uses for unfolding variants, we do not consider this operator further.

\subsection{Generic Programming for Non-Commutative Rows}
\label{sec:non-commutative}
\newcommand\nc[1]{{#1}_\mathsf{nc}}

\Rose encompasses multiple models of rows.  Our discussion so far has assumed \emph{simple} rows, a commutative row theory which seems to be the most natural approach to typing records and variants.  However, other theories may be more suited to particular applications.  For example, scoped rows~\citep{BerthomieuM95,Leijen05}, a non-commutative row theory, are particularly well suited to capturing algebraic effects and handlers.  A language that includes both extensible data types and algebraic effects, then, might want to include both simple rows (for data types) and scoped rows (for effects).  Alternatively, a language could support encoding algebraic effects via extensible data types, such as by using free monads.  But then, to capture effects naturally, the language could support extensible data types over both simple and scoped rows!


The challenge in adapting our account to non-commutative row theories is that we no longer have a single idea of containment.  The same label $\ell$, or indeed the same labeled type $\LabTy \ell \tau$, may appear multiple times in a single row.  To support non-commutative row theories, \Rose introduced two containment operators: the ``left'' version, $\Leqp[\Left] {\rho_1} {\rho_2}$, which holds if there is a $\rho'$ such that $\RowPlusP {\rho_1} {\rho'} {\rho_2}$, and the ``right'' version, $\Leqp[\Right] {\rho_1} {\rho_2}$, which holds if there is a $\rho'$ such that $\RowPlusP {\rho'} {\rho_1} {\rho_2}$.

Unfortunately, neither of these is a drop-in replacement for the predicates in our generic operators, as individual entries need not be at either the beginning or end of the input row.  We can apply a similar idea, by replacing the constraint $\RowPlusP {\Row {\LabTy l u}} y \rho$ with $\RowPlusP {\RowPlus {y_1} {\Row {\LabTy l u}}} {y_2} \rho$.  (Note that we cannot define a corresponding ``containment'' predicate: $l$, $u$, and $\rho$ do not uniquely determine $y_1$ and $y_2$.)  As we only have a binary row combination predicate, we express this by $\RowPlusP {y_1} {\Row {\LabTy l u}} z, \RowPlusP z {y_2} \rho$:
\begin{smalle}
\allowdisplaybreaks
\begin{gather*}  
\ib{\irule[\trule{${\nc{\Ana}}$}]
          {\begin{array}{@{}c@{}}
             {\KindJ \Gamma \rho {\RowK \kappa}}
             \isp
             {\KindJ \Gamma \phi {\kappa \to \TypeK}}
             \\
             {\TypeJ \Gamma M {\forall l \co \LabK, u \co \kappa, \shade{y_1, z, y_2 \co \RowK \kappa}. \, \shade{(\RowPlusP {y_1} {\Row {\LabTy l u}} z, \RowPlusP z {y_2} \rho)} \then \Sing l \to \phi \, u \to \tau}}
           \end{array}};
          {\TypeJ \Gamma {\Ana [\phi] \, M} {\Sigma (\phi \, \rho) \to \tau}}}
\\          
\ib{\irule[\trule{$\nc{\Syn}$}]
          {\begin{array}{@{}c@{}}
             {\KindJ \Gamma \rho {\RowK \kappa}}
             \isp
             {\KindJ \Gamma \phi {\kappa \to \TypeK}}
             \\
             {\TypeJ \Gamma M {\forall l \co \LabK, u \co \kappa, \shade{y_1, z, y_2 \co \RowK \kappa}. \, \shade{(\RowPlusP {y_1} {\Row {\LabTy l u}} z, \RowPlusP z {y_2} \rho)} \then \Sing l \to \phi \, u}}
           \end{array}};
          {\TypeJ \Gamma {\Syn [\phi] \, M} {\Pi (\phi \, \rho)}}}
\\
\ib{\irule[\trule{$\nc{\FoldP}$}]
          {\begin{array}{@{}c@{}}
             {M_1 : \forall l \co \LabK, t \co \TypeK, \shade{y_1, z, y_2 \co \RowK \kappa}. \, \shade{(\RowPlusP {y_1} {\Row {\LabTy l u}} z, \RowPlusP z {y_2} \rho)} \then \Sing l \to t \to \upsilon}
             \\
             {\TypeJ \Gamma {M_2} {\upsilon \to \upsilon \to \upsilon}}
             \isp
             {\TypeJ \Gamma {M_3} \upsilon}
             \isp
             {\TypeJ \Gamma N {\Pi \rho}}
           \end{array}};
          {\TypeJ \Gamma {\FoldP \, M_1 \, M_2 \, M_3 \, N} \upsilon}}
\end{gather*}
\end{smalle}
These rules generalize those previously presented: in a commutative theory, if $\RowPlusP {\RowPlus {y_1} {\Row {\LabTy l u}}} {y_2} \rho$, then there is a $y$ such that $\RowPlusP {\Row {\LabTy l u}} y \rho$, given by $\RowPlusP {y_1} {y_2} y$, and conversely.

\section{The \RO Calculus}
\label{sec:ro}

This section provides a formal description of the syntax and type system of \RO.  As in \citet{MorrisM19}, \RO is parameterized by a \emph{row theory}, giving the intended interpretation of rows.  A row theory $\Thy$ is a triple $\langle \vdash_\Thy, \equiv_\Thy, \Vdash_\Thy \rangle$, where
\begin{itemize}
\item $\vdash_\Thy$ is a kinding relation, capturing when rows are well-formed;
\item $\equiv_\Thy$ is an equivalence relation, identifying rows; and,
\item $\Vdash_\Thy$ is an entailment relation, giving the meaning of the $\lesssim$ and $\odot$ predicates.
\end{itemize}
We write \RO[\Thy] to indicate \RO instantiated with theory $\Thy$.  Our description of \RO syntax~\cref{sec:ro-syntax}, types~\cref{sec:ro-types}, and terms~\cref{sec:ro-terms} are all given generically over an arbitrary row theory $\Thy$.  We then provide three concrete row theories.  The \emph{minimal} row theory~\cref{sec:ro-minimal} captures labeled rows, but makes no commitment to when (non-singleton) rows are well-formed.  The examples in the previous section are all well-typed given only the minimal row theory.  We then describe the \emph{simple} row theory~\cref{sec:ro-simple}, which captures commutative \Remy-style rows, and the \emph{scoped} row theory~\cref{sec:ro-scoped}, which captures non-commutative Leijen-style rows.  We develop the expected metatheory in the following section, when we discuss our denotational interpretation of \RO in Agda.

\subsection{Syntax}
\label{sec:ro-syntax}

The syntax of \RO [\Thy] is given in \cref{fig:syntax}.  

\begin{figure}
\begin{smalle}
\begin{gather*}
\begin{array}{r@{\hspace{5px}}l@{\qquad}r@{\hspace{5px}}l@{\qquad}r@{\hspace{5px}}l@{\qquad}r@{\hspace{5px}}l}
  \text{Term variables} & x & \text{Type variables} & \alpha & \text{Labels} & \ell & \text{Directions} & d \in \Set {\Left, \Right}
\end{array}
\\[5px]
\begin{doublesyntaxarray}
  \mcl{\text{Kinds}} & \kappa & ::= & \TypeK \mid \LabK \mid \RowK \kappa \mid \kappa \to \kappa \\
  \mcl{\text{Predicates}} & \pi, \psi & ::= & \Leqp [d] \rho \rho \mid \RowPlusP \rho \rho \rho \\
  \text{Types} & \mcr{\phi, \tau, \upsilon, \rho, \xi} & ::= & \alpha \mid (\to) \mid \pi \then \tau \mid \forall \alpha\co\kappa. \tau \mid \lambda \alpha\co\kappa. \tau \mid \tau \, \tau \\
  & & & \mid & \ell \mid \Sing \xi \mid \LabTy \xi \tau \mid \Row {\tau_1, \dots, \tau_n} \mid \Pi \rho \mid \Sigma \rho \\
  \mcl{\text{Terms}} & M, N & ::= & x \mid \lambda x \co \tau. M \mid M \, N \mid \Lambda \alpha \co \kappa. M \mid \AppT M \tau \\
  & & & \mid & \ell \mid \LabTerm M M \mid \Unlabel M M \mid \Prj [d] M \mid M \Concat M \mid \Inj [d] M \mid M \Branch M \\
  & & & \mid & \Syn [\phi] \, M \mid \Ana [\phi]\, M \mid \FoldP \, M \, M \, M \, M \\
  \mcl{\text{Environments}} & \Gamma & ::= & \varepsilon \mid \Gamma, \alpha : \kappa \mid \Gamma, x : \tau \mid \Gamma, \pi
\end{doublesyntaxarray}
\end{gather*}
\end{smalle}
\caption{Syntax}
\label{fig:syntax}
\end{figure}

Kinds include types $\TypeK$, labels $\LabK$, rows $\RowK \kappa$ of kind $\kappa$, and type constructors $\kappa \to \kappa$.  Not all possible kinds are currently used in \RO. For example: while nothing prevents describing a type of kind $\RowK \LabK$ (i.e., a row of labels), we have no primitives that operate on such a type, and indeed suspect that such a type would be very difficult to use~\cref{sec:conclusion}.

Predicates include containment $\Leqp [d] \rho \rho$ and combination $\RowPlusP \rho \rho \rho$.  To account for non-commutative row theories, we include directed variants of the containment predicate; intuitively, if $\RowPlusP {\rho_1} {\rho_2} {\rho_3}$, then $\Leqp [\Left] {\rho_1} {\rho_3}$ and $\Leqp [\Right] {\rho_2} {\rho_3}$.  Given a commutative row theory, these predicates are equivalent. In a practical language based on \RO, we anticipate that the predicate language would be extended with other forms of predicates, such as type classes \citep{WadlerB89}, linearity constraints \citep{GanTM15,Morris16}, or general equality constraints \citep{PeytonJonesVWW06}.

We let $\phi, \tau, \upsilon, \rho$ and $\xi$ range over types; when possible, we use $\phi$ where we expect a type constructor, $\rho$ where we expect a row type, and $\xi$ where we expect a label.  Standard type constructs include variables $\alpha$, constants (here only the function arrow), quantifiers, abstractions, and applications.  Predicates appear in qualified types $\pi \then \tau$.  To incorporate labeling, we include labels ($\ell$) themselves, singletons $\Sing \xi$, and labeled types $\LabTy \xi \tau$.  Following \Rose, we treat labeled types and row types independently.  Finally, we include rows $\Row {\tau_1, \dots, \tau_n}$ (including the empty row), records, and variants.  Well-formedness of concrete rows is delegated to the row theory $\Thy$. 

We let $M,N$ range over terms.  Standard terms include variables, type and term abstractions, and applications.  Introduction and elimination of qualifiers is implicit.  To support labeling terms, we include label (singleton) constants $\ell$ and terms to label ($\LabTerm M M$) and unlabel ($\Unlabel M M$).  As the singleton record and variant types are isomorphic to their underlying single field or constructor type, we do not provide separate syntax to construct singleton records and variants.  Finally, we include the (directed) variant and record operators of \Rose, and the label-generic operators new to \RO [\Thy].

Environments track three kinds of assumptions: kindings of type variables $\alpha : \kappa$, typings of term variables $x : \tau$, and predicates $\pi$ (as qualified type elimination is implicit, we do not need to name predicate assumptions).  We combine these assumptions into a single context $\Gamma$ simply to avoid a superfluity of (mostly unchanging) metavariables.

\subsection{Types and Kinds}
\label{sec:ro-types}
 
\cref{fig:kinding} gives rules for context formation ($\EnvJ \Gamma$), kinding ($\KindJ \Gamma \tau \kappa$), and predicate formation ($\PredJ \Gamma \pi$), parameterized by row theory $\Thy$.

\begin{figure}
\small
\begin{gather*}
\fbox{$\EnvJ \Gamma$}
\\
\ib{\irule[\crule{emp}]{ };{\EnvJ \varepsilon}}
\rsp
\ib{\irule[\crule{tvar}]
          {\EnvJ \Gamma};
          {\EnvJ {\Gamma, \alpha : \kappa}}}
\rsp
\ib{\irule[\crule{var}]
          {\EnvJ \Gamma}
          {\KindJ \Gamma \tau \TypeK};
          {\EnvJ {\Gamma, x : \tau}}}
\rsp
\ib{\irule[\crule{pred}]
          {\EnvJ \Gamma}
          {\PredJ \Gamma \pi};
          {\EnvJ {\Gamma, \pi}}}
\end{gather*}
\begin{gather*}
\fbox{$\KindJ \Gamma \tau \kappa$} \; \fbox{$\PredJ \Gamma \pi$}
\\
\ib{\irule[\krule{var}]
          {\EnvJ \Gamma}
          {\alpha : \kappa \in \Gamma};
          {\Gamma \vdash \alpha : \kappa}}
\rsp
\ib{\irule[\krule{$(\to)$}]
          {\EnvJ \Gamma};
          {\KindJ \Gamma {(\to)} {\TypeK \to \TypeK \to \TypeK}}}
\rsp
\ib{\irule[\krule{$\then$}]
          {\PredJ \Gamma \pi}
          {\KindJ {\Gamma, \pi} \tau \TypeK};
          {\KindJ \Gamma {\pi \then \tau} \TypeK}}
\\
\ib{\irule[\krule{$\forall$}]
          {\KindJ {\Gamma, \alpha : \kappa} \tau \TypeK};
          {\KindJ \Gamma {\forall \alpha\co\kappa. \tau} \TypeK}}
\rsp
\ib{\irule[\krule{$\I\to$}]
          {\KindJ {\Gamma, \alpha : \kappa_1} \tau \kappa_2};
          {\KindJ \Gamma {\lambda \alpha \co \kappa_1. \tau} {\kappa_1 \to \kappa_2}}}
\rsp
\ib{\irule[\krule{$\E\to$}]
          {\KindJ \Gamma {\tau_1} {\kappa_1 \to \kappa_2}}
          {\KindJ \Gamma {\tau_2} {\kappa_1}};
          {\KindJ \Gamma {\tau_1 \, \tau_2} {\kappa_2}}}
\\
\ib{\irule[\krule{lab}]
          {\EnvJ \Gamma};
          {\KindJ \Gamma \ell \LabK}}
\rsp
\ib{\irule[\krule{sing}]
          {\KindJ \Gamma \xi \LabK};
          {\KindJ \Gamma {\Sing\xi} \TypeK}}
\rsp
\ib{\irule[\krule{lty}]
          {\KindJ \Gamma \xi \LabK}
          {\KindJ \Gamma \tau \kappa};
          {\KindJ \Gamma {\LabTy \xi \tau} \kappa}}
\rsp
\ib{\irule[\krule{row}]
          {\KindJS \Thy \Gamma {\Row {\overline {\LabTy \xi \tau}}} {\RowK \kappa}};
          {\KindJ \Gamma {\Row{\overline {\LabTy \xi \tau}}} {\RowK\kappa}}}
\\
\ib{\irule[\krule{$\Pi$}]
          {\KindJ \Gamma \rho {\RowK \kappa}};
          {\KindJ \Gamma {\Pi\rho} \kappa}}
\rsp
\ib{\irule[\krule{$\Sigma$}]
          {\KindJ \Gamma \rho {\RowK \kappa}};
          {\KindJ \Gamma {\Sigma\rho} \kappa}}
\rsp
\ib{\irule[\krule{lift$_1$}]
          {\KindJ \Gamma \rho {\RowK{\kappa_1 \to \kappa_2}}}
          {\KindJ \Gamma \tau {\kappa_1}};
          {\KindJ \Gamma {\rho\,\tau} {\RowK{\kappa_2}}}}
\\
\ib{\irule[\krule{lift$_2$}]
          {\KindJ \Gamma \phi {\kappa_1 \to \kappa_2}}
          {\KindJ \Gamma \rho {\RowK{\kappa_1}}};
          {\KindJ \Gamma {\phi\,\rho} {\RowK{\kappa_2}}}}
\rsp          
\ib{\irule[\krule{$\lesssim_{d}$}]
          {\KindJ \Gamma {\rho_i} {\RowK \kappa}};
          {\PredJ \Gamma {\Leqp [d] {\rho_1} {\rho_2}}}}
\rsp
\ib{\irule[\krule{$\odot$}]
          {\KindJ \Gamma {\rho_i} {\RowK \kappa}};
          {\PredJ \Gamma {\RowPlusP {\rho_1} {\rho_2} {\rho_3}}}}
\end{gather*}

\caption{Contexts and kinding.}
\label{fig:kinding}
\end{figure}

The kinding rules are mostly standard.  Rule \krule{row} delegates well-formedness of rows to the row theory $\Thy$.  Rules \krule{$\Pi$} and \krule{$\Sigma$} capture the formation of record and variant types, lifted to arbitrary kinds $\kappa$.  In the functor example~(\secfig{maps}{functors}), we had a row of type constructors $z : \RowK {\TypeK \to \TypeK}$.  Applying \krule{$\Sigma$}, we can conclude that $\Sigma z : \TypeK \to \TypeK$, and so that $(\Sigma z) \, t : \TypeK$.


Rules \krule{lift$_1$} and \krule{lift$_2$} license the lifting that has played a prominent role in our examples.  Rule \krule{lift$_2$} says that a type constructor $\kappa_1 \to \kappa_2$, applied to a row of $\kappa_1$s, generates a row of $\kappa_2$s.  Consider a type like $z \to t$, or more pedantically $(\to) \, z \, t$, where $z : \RowK \TypeK$ and $t : \TypeK$.  We begin by applying \krule{lift$_2$} to apply $(\to)$ to $z$, concluding $(\to) \, z : \RowK {\TypeK \to \TypeK}$.  Then, we apply \krule{lift$_1$} to apply $(\to) \, z$ to $t$, concluding that $(\to) \, z \, t : \RowK \TypeK$.

One might argue that these are simply syntactic abbreviations, and complicate the reading of types.  Instead, we should follow the lead of Featherweight Ur \citep{Chlipala10}, and use an explicit $\mathsf{map}$ operation to lift types over rows.  However, in developing our examples, we found that the extra weight introduced by a more explicit approach obscured the meaning of the terms.  For example, contrast our types for $reify$ and $reflect$~(\secfig{reify-reflect}{reify-reflect}) with the more explicit
\begin{align*}
  \ExC{reify} &: \forall z \co \RowK\TypeK, t : \TypeK. \, (\Sigma z \to t) \to \Pi (\mathsf{map} \, (\lambda (s \co \TypeK). \, s \to t) \, z) \\
  \ExC{reflect} &: \forall z \co \RowK\TypeK, t : \TypeK. \, \Pi (\mathsf{map} \, (\lambda (s \co \TypeK). \, s \to t) \, z) \to (\Sigma z \to t)
\end{align*}
Or, similarly, contrast our type for $\FmapS$~(\secfig{maps}{functors}) with the more explicit
\begin{align*}
  \FmapS &: \forall z \co \RowK {\TypeK \to \TypeK}. \, \Pi (\mathsf{map} \, \TyC{Functor} \, z) \to \TyC{Functor} \, (\lambda (t \co \TypeK). \, \Sigma \, (\mathsf {map} \, (\lambda f \co \TypeK \to \TypeK. \, f \, t) \, z))
\end{align*}
But in the end, this is a matter of taste; restricting $\Sigma$ and $\Pi$ to arguments of kind $\RowK \TypeK$ and making row mapping explicit would not fundamentally restrict the expressiveness of \RO.

\begin{figure}
\begin{gather*}
\fbox{$\EqvJ \tau \tau$} \; \fbox{$\EqvJ \pi \pi$}
\\
\ib{\irule[\erule{refl}]
          { };
          {\EqvJ \tau \tau}}
\rsp
\ib{\irule[\erule{sym}]
          {\EqvJ {\tau_1} {\tau_2}};
          {\EqvJ {\tau_2} {\tau_1}}}
\rsp
\ib{\irule[\erule{trans}]
          {\EqvJ {\tau_1} {\tau_2}}
          {\EqvJ {\tau_2} {\tau_3}};
          {\EqvJ {\tau_1} {\tau_3}}}
\rsp
\ib{\irule[\erule{$\beta$}]
          { };
          {\EqvJ {(\lambda \alpha\co\kappa. \tau)\,\upsilon} {\tau[\upsilon/\alpha]}}}
\\
\ib{\irule[\erulec {\then}]
          {\EqvJ {\pi_1} {\pi_2}}
          {\EqvJ {\tau_1} {\tau_2}};
          {\EqvJ {\pi_1 \then \tau_1} {\pi_2 \then \tau_2}}}
\rsp
\ib{\irule[\erulec{\forall}]
          {\EqvJ {\tau[\gamma/\alpha]} {\upsilon[\gamma/\beta]}};
          {\EqvJ {\forall\alpha\co\kappa.\tau} {\forall\beta\co\kappa.\upsilon}}!
          {\gamma\not\in fv(\tau, \upsilon)}}
\rsp
\ib{\irule[\erulec{\text{\textsc{app}}}]
          {\EqvJ {\tau_i} {\upsilon_i}};
          {\EqvJ {\tau_1\,\tau_2} {\upsilon_1\,\upsilon_2}}}
\\
\ib{\irule[\erulec{\triangleright}]
          {\EqvJ {\xi_1} {\xi_2}}
          {\EqvJ {\tau_1} {\tau_2}};
          {\EqvJ {\LabTy {\xi_1} {\tau_1}} {\LabTy {\xi_2} {\tau_2}}}}
\rsp
\ib{\irule[\erule{row}]
          {\EqvJS \Thy {\Row {\overline {\LabTy {\xi_i} {\tau_i}}}} {\Row {\overline {\LabTy {\xi'_j} {\tau'_j}}}}};
          {\EqvJ {\Row {\overline {\LabTy {\xi_i} {\tau_i}}}} {\Row {\overline {\LabTy {\xi'_j} {\tau'_j}}}}}}
\rsp
\ib{\irule[\erulec{\Sing\cdot}]
          {\EqvJ {\xi_1} {\xi_2}};
          {\EqvJ {\Sing{\xi_1}} {\Sing{\xi_2}}}}
\\
\ib{\irule[\erule{lift$_1$}]
          { };
          {\EqvJ {\Row{\LabTy \xi \phi } \, \tau} {\Row{\LabTy {\xi} {\phi\,\tau}}}}}
\rsp
\ib{\irule[\erule{lift$_2$}]
          { };
          {\EqvJ {\phi \, \Row{\LabTy {\xi} {\tau}}} {\Row{\LabTy {\xi} {\phi\,\tau}}}}}
\rsp
\\
\ib{\irule[\erulec{\Pi\Sigma}]
          {\EqvJ {\rho_1} {\rho_2}};
          {\EqvJ {K \rho_1} {K \rho_2}}}
\rsp
\ib{\irule[\erule{lift$_3$}]
          { };
          {\EqvJ {(K \rho) \, \tau} {K (\rho \, \tau)}}}
\rsp
\ib{\irule[\erule{sing}]
          { };
          {\EqvJ {K \Row{\LabTy \xi \tau}} {\LabTy \xi \tau}}}
\rsp
(K \in \Set {\Pi, \Sigma})
\\
\ib{\irule[\erulec{\lesssim_{d}}]
          {\EqvJ {\tau_i} {\upsilon_i}};
          {\EqvJ {\Leqp [d] {\tau_1} {\tau_2}} {\Leqp [d] {\upsilon_1} {\upsilon_2}}}}
\rsp
\ib{\irule[\erulec{\odot}]
          {\EqvJ {\tau_i} {\upsilon_i}};
          {\EqvJ {\RowPlusP {\tau_1} {\tau_2} {\tau_3}} {\RowPlusP {\upsilon_1} {\upsilon_2} {\upsilon_3}}}}
\end{gather*}
\caption{Type and predicate equivalence}
\label{fig:type-equiv}
\end{figure}

The type equivalence rules are shown in \cref{fig:type-equiv}.  The first three lines are standard.  The rules \erule{lift$_1$} and \erule{lift$_2$} realize the promise made in \krule{lift$_1$} and \krule{lift$_2$}, lifting single type operators or type arguments to rows.  Rule \erule{row} delegates equivalence of row types to the row theory $\Thy$.  Rule \erule{lift$_3$} gives $\Pi$ and $\Sigma$ their intended meaning at higher kinds.  Finally, \erule{sing} captures the isomorphism between singleton records, singleton variants, and their underlying field (or constructor) type.  Again, this latter rule is not integral to \RO; a more explicit version, with separate terms to introduce and eliminate singleton records and variants, would be just as expressive.

\subsection{Terms}
\label{sec:ro-terms}

\renewcommand\EntJ[2]{\EntJS \Thy #1 #2}
\begin{figure}
\small  
\begin{gather*}
\fbox{$\TypeJ \Gamma M \tau$}
\\
\ib{\irule[\trule{var}]
          {\EnvJ \Gamma}
          {x : \tau \in \Gamma};
          {\TypeJ \Gamma x \tau}}
\rsp
\ib{\irule[\trule{$\I\to$}]
          {\KindJ \Gamma {\tau_1} {\TypeK}}
          {\TypeJ {\Gamma, x : \tau_1} M {\tau_2}};
          {\TypeJ \Gamma {\lambda x : \tau_1. M} {\tau_1 \to \tau_2}}}
\rsp          
\ib{\irule[\trule{$\E\to$}]
          {\TypeJ \Gamma {M_1} {\tau_1 \to \tau_2}}
          {\TypeJ \Gamma {M_2} {\tau_1}};
          {\TypeJ \Gamma {M_1 \, M_2} {\tau_2}}}
\\
\ib{\irule[\trule{$\equiv$}]
          {\TypeJ \Gamma M \tau}
          {\EqvJ \tau \upsilon};
          {\TypeJ \Gamma M \upsilon}}
\rsp
\ib{\irule[\trule{$\I\then$}]
          {\PredJ \Gamma \pi}
          {\TypeJ {\Gamma, \pi} M \tau};
          {\TypeJ \Gamma M {\pi \then \tau}}}
\rsp
\ib{\irule[\trule{$\E\then$}]
          {\TypeJ \Gamma M {\pi \then \tau}}
          {\EntJ \Gamma \pi};
          {\TypeJ \Gamma M \tau}}
\\
\ib{\irule[\trule{$\I\forall$}]
          {\TypeJ {\Gamma, \alpha : \kappa} M \tau};
          {\TypeJ \Gamma {\Lambda \alpha \co \kappa. M} {\forall \alpha\co\kappa. \tau}}}
\rsp
\ib{\irule[\trule{$\E\forall$}]
          {\TypeJ \Gamma M {\forall \alpha\co\kappa. \tau}}
          {\KindJ \Gamma \upsilon \kappa};
          {\TypeJ \Gamma {\AppT M \upsilon} {\tau[\upsilon/\alpha]}}}
\\
\ib{\irule[\trule{sing}]
          {\EnvJ \Gamma};
          {\TypeJ \Gamma \ell {\Sing \ell}}}
\rsp
\ib{\irule[\trule{$\I\triangleright$}]
          {\TypeJ \Gamma {M_1} {\Sing \ell}}
          {\TypeJ \Gamma {M_2} \tau};
          {\TypeJ \Gamma {\LabTerm {M_1} {M_2}} {\LabTy \ell \tau}}}
\rsp
\ib{\irule[\trule{$\E\triangleright$}]
          {\TypeJ \Gamma {M_1} {\LabTy \ell \tau}}
          {\TypeJ \Gamma {M_2} {\Sing \ell}};
          {\TypeJ \Gamma {\Unlabel {M_1} {M_2}} \tau}}
\\
\ib{\irule[\trule{$\E\Pi$}]
          {\TypeJ \Gamma M {\Pi \rho_1}}
          {\EntJ \Gamma {\Leqp [d] {\rho_2} {\rho_1}}};
          {\TypeJ \Gamma {\Prj [d] M} {\Pi \rho_2}}}
\rsp
\ib{\irule[\trule{$\I\Pi$}]
          {\TypeJ \Gamma {M_1} {\Pi \rho_1}}
          {\TypeJ \Gamma {M_2} {\Pi \rho_2}}
          {\EntJ \Gamma {\RowPlusP {\rho_1} {\rho_2} {\rho_3}}};
          {\TypeJ \Gamma {M_1 \Concat M_2} {\Pi \rho_3}}}
\\
\ib{\irule[\trule{$\I\Sigma$}]
          {\TypeJ \Gamma M {\Sigma \rho_1}}
          {\EntJ \Gamma {\Leqp [d] {\rho_1} {\rho_2}}};
          {\TypeJ \Gamma {\Inj [d] M} {\Sigma \rho_2}}}
\rsp
\ib{\irule[\trule{$\E\Sigma$}]
          {\TypeJ \Gamma {M_1} {\Sigma \rho_1 \to \tau}}
          {\TypeJ \Gamma {M_2} {\Sigma \rho_2 \to \tau}}
          {\EntJ \Gamma {\RowPlusP {\rho_1} {\rho_2} {\rho_3}}};
          {\TypeJ \Gamma {M_1 \Branch M_2} {\Sigma \rho_3 \to \tau}}}
\\
\ib{\irule[\trule{$\Ana$}]
          {\begin{array}{@{}c@{}}
             {\KindJ \Gamma \rho {\RowK \kappa}}
             \isp
             {\KindJ \Gamma \phi {\kappa \to \TypeK}}
             \\
             {\TypeJ \Gamma M {\forall l \co \LabK, u \co \kappa, y_1, z, y_2 \co \RowK \kappa. \, (\RowPlusP {y_1} {\Row {\LabTy l u}} z, \RowPlusP z {y_2} \rho) \then \Sing l \to \phi \, u \to \tau}}
           \end{array}};
          {\TypeJ \Gamma {\Ana [\phi] \, M} {\Sigma (\phi \, \rho) \to \tau}}}
\\          
\ib{\irule[\trule{$\Syn$}]
          {\begin{array}{@{}c@{}}
             {\KindJ \Gamma \rho {\RowK \kappa}}
             \isp
             {\KindJ \Gamma \phi {\kappa \to \TypeK}}
             \\
             {\TypeJ \Gamma M {\forall l \co \LabK, u \co \kappa, y_1, z, y_2 \co \RowK \kappa. \, (\RowPlusP {y_1} {\Row {\LabTy l u}} z, \RowPlusP z {y_2} \rho) \then \Sing l \to \phi \, u}}
           \end{array}};
          {\TypeJ \Gamma {\Syn [\phi] \, M} {\Pi (\phi \, \rho)}}}
\\
\ib{\irule[\trule{$\FoldP$}]
          {\begin{array}{@{}c@{}}
             {M_1 : \forall l \co \LabK, t \co \TypeK, y_1, z, y_2 \co \RowK \kappa. \, (\RowPlusP {y_1} {\Row {\LabTy l u}} z, \RowPlusP z {y_2} \rho) \then \Sing l \to t \to \upsilon}
             \\
             {\TypeJ \Gamma {M_2} {\upsilon \to \upsilon \to \upsilon}}
             \isp
             {\TypeJ \Gamma {M_3} \upsilon}
             \isp
             {\TypeJ \Gamma N {\Pi \rho}}
           \end{array}};
          {\TypeJ \Gamma {\FoldP \, M_1 \, M_2 \, M_3 \, N} \upsilon}}
\end{gather*}
\caption{Typing}
\label{fig:typing}
\end{figure}
\renewcommand\EntJ[2]{#1 \Vdash #2}

\noindent
\cref{fig:typing} gives the typing rules for \RO. We have already developed its novelties in the previous section, but will briefly highlight the remaining features of the type system.  Lines 1--3 contain a standard treatment of functions, qualified types, and quantified types.  Rule \trule{sing} is used to introduce label singleton constants, which can then be used to label \trule{\I\triangleright} or unlabel \trule{\E\triangleright} terms.  Rule \trule{$\equiv$} can be used (among other things) to move between labeled terms and singleton records or variants.  The rules for projection, concatenation, injection, and branching are identical to the corresponding rules for \Rose.  Finally, the rules for analyzing variants and synthesizing and folding rows are discussed in the previous section.

\subsection{Minimal Rows}
\label{sec:ro-minimal}

\cref{fig:minimal} gives the minimal row theory $\mathcal M$.

\begin{figure}
\renewcommand\EntJ[2]{\EntJS \Mty {#1} {#2}}
\small
\begin{gather*}
\fbox{$\KindJS \Mty \Gamma \rho \kappa$} \; \fbox{$\vphantom\Gamma \EqvJS \Mty \rho \rho$}
\\
\ib{\irule[\krule{mrow}]
          {\KindJ \Gamma \xi \LabK}
          {\KindJ \Gamma \tau \kappa};
          {\KindJS \Mty \Gamma {\Row {\LabTy \xi \tau}} {\RowK \kappa}}}
\rsp
\ib{\irule[\erule{mrow}]
          {\EqvJ \xi {\xi'}}
          {\EqvJ \tau {\tau'}};
          {\EqvJS \Mty {\Row {\LabTy \xi \tau}} {\Row {\LabTy {\xi'} {\tau'}}}}}
\\          
\fbox{$\EntJ \Gamma \pi$}
\\
\ib{\irule[\entrule{ax}]
          {\pi \in \Gamma};
          {\EntJ {\Gamma} \pi}}
\rsp
\ib{\irule[\entrule{refl}]
          { };
          {\EntJ \Gamma {\Leqp [d] \rho \rho}}}
\rsp
\ib{\irule[\entrule{trans}]
          {\EntJ \Gamma {\Leqp [d] {\rho_1} {\rho_2}}}
          {\EntJ \Gamma {\Leqp [d] {\rho_2} {\rho_3}}};
          {\EntJ \Gamma {\Leqp [d] {\rho_1} {\rho_3}}}}
\\
\ib{\irule[\entrule{$\equiv$}]
          {\EntJ \Gamma {\pi_1}}
          {\pi_1 \equiv \pi_2};
          {\EntJ \Gamma {\pi_2}}}
\rsp
\ib{\irule[\entrule{$\lesssim$lift$_1$}]
          {\EntJ \Gamma {\Leqp [d] {\rho_1} {\rho_2}}};
          {\EntJ \Gamma {\Leqp [d] {\phi\,\rho_1} {\phi\,\rho_2}}}}
\rsp
\ib{\irule[\entrule{$\lesssim$lift$_2$}]
          {\EntJ \Gamma {\Leqp [d] {\rho_1} {\rho_2}}};
          {\EntJ \Gamma {\Leqp [d] {\rho_1\,\tau} {\rho_2\,\tau}}}}
\\
\ib{\irule[\entrule{$\odot$lift$_1$}]
          {\EntJ \Gamma {\RowPlusP {\rho_1} {\rho_2} {\rho_3}}};
          {\EntJ \Gamma {\RowPlusP {\rho_1 \, \tau} {\rho_2 \, \tau} {\rho_3 \, \tau}}}}
\rsp
\ib{\irule[\entrule{$\odot$lift$_2$}]
          {\EntJ \Gamma {\RowPlusP {\rho_1} {\rho_2} {\rho_3}}};
          {\EntJ \Gamma {\RowPlusP {\phi\,\rho_1} {\phi\,\rho_2} {\phi\,\rho_3}}}}
\\
\ib{\irule[\entrule{${\odot}{\lesssim_\Left}$}]
          {\EntJ \Gamma {\RowPlusP {\rho_1} {\rho_2} {\rho_3}}};
          {\EntJ \Gamma {\Leqp [\Left] {\rho_1} {\rho_3}}}}
\rsp
\ib{\irule[\entrule{${\odot}{\lesssim_\Right}$}]
          {\EntJ \Gamma {\RowPlusP {\rho_1} {\rho_2} {\rho_3}}};
          {\EntJ \Gamma {\Leqp [\Right] {\rho_2} {\rho_3}}}}
\end{gather*}
\caption{Minimal row theory $\mathcal M = \langle \vdash_\Mty, \equiv_\Mty, \Vdash_{\Mty} \rangle$}
\label{fig:minimal}
\end{figure}

The minimal row theory only includes singleton rows, and so \RO [\mathcal M] can express very few practical uses of extensible data types.  However, the minimal row theory captures the fundamental properties that all (labeled) row theories share.  Our motivating examples~\cref{sec:generic} all type in \RO [\mathcal M].

The interesting content of the minimal row theory is its entailment relation.  Rules \entrule{refl} and \entrule{trans} make containment a preorder.  Rules \entrule{$\lesssim$lift$_1$} and \entrule{$\lesssim$lift$_2$} capture that containment is preserved by lifted application.  Rules \entrule{$\odot$lift$_1$} and \entrule{$\odot$lift$_2$} similarly capture that combination is preserved by lifted application.  Finally, rules \entrule{${\odot}{\lesssim}$l} and \entrule{${\odot}{\lesssim}$r} capture the relationship between containment and combination.

\subsection{Simple Rows}
\label{sec:ro-simple}

\newcommand\One{}
\newcommand\Two{'}
\newcommand\Three{''}
\newcommand\Apart{\mathbin{\#}}
\newcommand\RowIx[4]{\Rowlr{\overline{#4}}^{\!#1 \in {#2} \dots {#3}}}

\cref{fig:simple} gives the simple row theory $\mathcal S$.

\begin{figure}
\small  
\begin{gather*}
\fbox{$\KindJS \Sty \Gamma \rho {\RowK\kappa}$} \; \fbox{$\vphantom{\Gamma}\EqvJS \Sty \rho \rho$}
\\
\begin{gathered}
\ib{\irule[\krule{srow}]
          {\KindJ \Gamma {\xi_i} \LabK}
          {\KindJ \Gamma {\tau_i} \kappa}
          {\forall i, j \not= i. \, \xi_i \Apart \xi_j};
          {\KindJS \Sty \Gamma {\RowIx i 1 n {\LabTy {\xi_i} {\tau_i}}} {\RowK \kappa}}}
\\
\text{where $\xi \Apart \xi'$ iff $\xi = \ell, \xi' = \ell', \ell \not= \ell'$}
\end{gathered}
\rsp
\begin{gathered}
\ib{\irule[\erule {srow}]
          {\EqvJ {\xi\One_i} {\xi\Two_{p(i)}}}
          {\EqvJ {\tau\One_i} {\tau\Two_{p(i)}}};
          {\EqvJS \Sty {\RowIx i 1 n {\LabTy {\xi\One_i} {\tau\One_i}}} {\RowIx j 1 n {\LabTy {\xi\Two_j} {\tau\Two_j}}}}}
\\
\text{where $p$ permutes $1 \dots n$}          
\end{gathered}
\\
\fbox{$\EntJS \Sty \Gamma \pi$}
\\
\text{(the rules of $\Vdash_\Mty$)}
\rsp
\begin{gathered}
\ib{\irule[\entsrule \Sty {$\lesssim_{d}$}]
          {\EqvJ {\xi\One_i} {\xi\Two_{p(i)}}}
          {\EqvJ {\tau\One_i} {\tau\Two_{p(i)}}};
          {\EntJS \Sty \Gamma {\Leqp [d] {\RowIx i 1 m {\LabTy {\xi\One_i} {\tau\One_i}}} {\RowIx j 1 n {\LabTy {\xi\Two_j} {\tau\Two_j}}}}}}
\\
\text{where $p$ injects $1 \dots m$ into $1 \dots n$}          
\end{gathered}
\\
\ib{\irule[\entsrule \Sty {$\odot$}]
          {\EqvJ {\xi\One_i} {\xi\Three_{p(i)}}}
          {\EqvJ {\tau\One_i} {\tau\Three_{p(i)}}}
          {\EqvJ {\xi\Two_j} {\xi\Three_{r(j)}}}
          {\EqvJ {\tau\Two_j} {\tau\Three_{r(j)}}};
          {\EntJS \Sty \Gamma {\RowPlusP {\RowIx i 1 m {\LabTy {\xi\One_i} {\tau\One_i}}} {\RowIx j 1 n {\LabTy {\xi\Two_j} {\tau\Two_j}}} {\RowIx k 1 {m + n} {\LabTy {\xi\Three_k} {\tau\Three_k}}}}}}
\\
\text{where $p$ injects $1 \dots m$ into $1 \dots m + n$, $r$ injects $1 \dots n$ into $1 \dots m + n$, and for all $i, j$, $p(i) \not= r(j)$}
\end{gather*}
\caption{The simple row theory $\mathcal S = \langle \vdash_\Sty, \equiv_\Sty, \Vdash_\Sty \rangle$; the entailment relation extends $\Vdash_\Mty$.}
\label{fig:simple}
\end{figure}

The simple theory is a commutative theory, in which labels may appear at most once in any row; it captures the most common approach to row types, originally introduced by \citet{Remy89}. The challenge to expressing the simple row theory in \RO arises from first-class labels.  As noted by \citet{Leijen04}, among others, first-class labels can introduce surprising corner cases.  Consider a type like $\Pi \Row {\LabTy {\xi_1} {Int}, \LabTy {\xi_2} {Int}}$, where $\xi_1$ and $\xi_2$ are types of kind $\LabK$.  This type only makes sense if $\xi_1$ and $\xi_2$ are guaranteed to be different labels.  This restriction is captured in \krule{srow}: each pair of labels in a row must be different concrete labels.  Of course, this condition is satisfied trivially for the empty and singleton rows.  Nor does this requirement limit the use of first-class labels, as longer rows may always be expressed as concatenations of singleton rows---indeed, such an elaboration could be done automatically, treating rows as partial type constructors~\citep{JonesD08,JonesME20,IngleHM22}.



The entailment relation extends that of the minimal row theory with rules for concrete rows.  In each case, the essential evidence is a mapping between rows in the predicate; as we will see in the next section, these mappings are exactly the information needed to implement the record and variant operations.  There are more generic entailment rules that could be useful in a practical realization of \RO [\Sty].  For example, combination gives a least upper bound for the containment relation:
\[
\ib{\irule{\EntJS \Mty \Gamma {\Leqp [d] {\rho_1} \rho}}
          {\EntJS \Mty \Gamma {\Leqp [d] {\rho_2} \rho}}
          {\EntJS \Mty \Gamma {\RowPlusP {\rho_1} {\rho_2} {\rho_3}}};
          {\EntJS \Mty \Gamma {\Leqp [d] {\rho_3} \rho}}}
\]
Nevertheless, the rules we give here capture the essential properties of the simple row theory; we regard further extension of the entailment relation as an orthogonal concern.




\subsection{Scoped Rows}
\label{sec:ro-scoped}

\cref{fig:scoped} gives the scoped row theory $\mathcal C$.

\begin{figure}
\small
\begin{gather*}
\fbox{$\KindJS \Cty \Gamma \rho {\RowK \kappa}$} \; \fbox{$\vphantom\Gamma \EqvJS \Cty \rho \rho$}
\\
\ib{\irule[\krule{crow}]
          {\KindJ \Gamma {\xi_i} \LabK}
          {\KindJ \Gamma {\tau_i} \kappa};
          {\KindJS \Cty \Gamma {\RowIx i 1 n {\LabTy {\xi_i} {\tau_i}}} {\RowK\kappa}}}
\rsp
\begin{gathered}
\ib{\irule[\erule{crow}]
          {\EqvJ {\xi_i} {\xi'_{p(i)}}}
          {\EqvJ {\tau_i} {\tau'_{p(i)}}};
          {\EqvJS \Cty {\RowIx i 1 n {\LabTy {\xi_i} {\tau_i}}} {\RowIx j 1 n {\LabTy {\xi'_j} {\tau'_j}}}}}
\\
\text{where $p$ permutes $1 \dots n$, if $i < j, p(i) > p(j)$, then $\xi_i \Apart \xi_j$}
\end{gathered}
\\
\fbox{$\EntJS \Cty \Gamma \pi$}
\\
\begin{gathered}
\text{(the rules of $\Vdash_\Mty$)}
\rsp
\ib{\irule[\entsrule \Cty {$\lesssim_\Left$}]
          {\EqvJ {\xi\One_i} {\xi\Two_j}}
          {\EqvJ {\tau\One_i} {\tau\Two_j}}
          {\text{for $i \in 1 \dots m, p(j) = i$}};
          {\EntJS \Cty \Gamma {\Leqp [\Left] {\RowIx i 1 m {\LabTy {\xi\One_i} {\tau\One_i}}} {\RowIx j 1 n {\LabTy {\xi\Two_j} {\tau\Two_j}}}}}}
\\
\text{where $p$ permutes $1 \dots n$, if $i < j$ and $p(i) > p(j)$, then $\xi\Two_i \Apart \xi\Two_j$}
\end{gathered}
\\
\begin{gathered}
\ib{\irule[\entsrule \Cty {$\lesssim_\Right$}]
          {\EqvJ {\xi\One_i} {\xi\Two_j}}
          {\EqvJ {\tau\One_i} {\tau\Two_j}}
          {\text{for $i \in 1 \dots m, p(j) = n - m + i$}};
          {\EntJS \Cty \Gamma {\Leqp [\Right] {\RowIx i 1 m {\LabTy {\xi\One_i} {\tau\One_i}}} {\RowIx j 1 n {\LabTy {\xi\Two_j} {\tau\Two_j}}}}}}
\\
\text{where $p$ permutes $1 \dots n$, if $i < j$ and $p(i) > p(j)$, then $\xi\Two_i \Apart \xi\Two_j$}
\end{gathered}
\\
\ib{\irule[\entsrule \Cty {$\odot$}]
          {\begin{array}{@{}c@{}}
            {\EqvJ {\xi\One_i} {\xi\Three_k}}
            \isp
            {\EqvJ {\tau\One_i} {\tau\Three_k}}
            \isp
            \text{for $i \in 1 \dots m, p(k) = i$}
            \\            
            {\EqvJ {\xi\Two_j} {\xi\Three_k}}
            \isp
            {\EqvJ {\tau\Two_j} {\tau\Three_k}}
            \isp
            \text{for $j \in 1 \dots n, p(k) = m + j$}
           \end{array}};
          {\EntJS \Cty \Gamma {\RowPlusP {\RowIx i 1 m {\LabTy {\xi\One_i} {\tau\One_i}}} {\RowIx j 1 n {\LabTy {\xi\Two_j} {\tau\Two_j}}} {\RowIx k 1 {m + n} {\LabTy {\xi\Three_k} {\tau\Three_k}}}}}}
\\
\text{where $p$ permutes $1 \dots m + n$, if $i < j$ and $p(i) > p(j)$ then $\xi\Three_i \Apart \xi\Three_j$.}
\end{gather*}

\caption{Scoped rows: kinding, entailment, and equivalence}
\label{fig:scoped}
\end{figure}

The scoped row theory is a non-commutative theory, in which the left-most instance of a given label is preferred; it was introduced by \citet{BerthomieuM95} and independently by \citet{Leijen05}.  Because labels can be repeated, there is no difficulty in the kinding rule \krule{crow}.  However, more care must be taken in the entailment relation: we want to allow $\Leqp [\Left] {\Row {\LabTy {\Lab y} {Int}}} {\Row {\LabTy {\Lab x} {Int}, \LabTy {\Lab y} {Int}}}$, as there is no harm in permuting distinct labels, while excluding $\Leqp [\Left] {\Row {\LabTy {\Lab x} {Bool}}} {\Row {\LabTy {\Lab x} {Int}, {\LabTy {\Lab x} {Bool}}}}$, as this permutes identical labels.  This is captured by the side condition on the permutations in each of the entailment rules, which requires that swapped labels be provably distinct.

\section{Interpreting (Stratified) \RO in Agda}
\label{sec:semantics}

We have two goals in defining semantics for \RO.  Primarily, of course, is to demonstrate the soundness of \RO's type system.  Secondarily  is to show that \RO need not introduce runtime dependence on or manipulation of labels compared to extensible data types without label-generic operators.

To accomplish both goals, we embedded \RO [\mathcal{M}] typings in the Agda type theory, and then defined a denotational interpretation of those typings in Agda itself, interpreting the \RO function space as Agda functions, \RO records and variants as dependent products and sums with finite natural indices, evidence for containment and combination as maps between finite naturals, and so forth.  In particular, labels are interpreted as the unit 
type, and the indexing of products and sums does not depend on the identities of labels in the source derivations.  

While our mechanization of the entailment relation is limited to the minimal row theory, our denotations are not correspondingly limited to singleton rows, records, and variants.  To the contrary, because our denotations do not depend on labels directly, they are sufficient for all the row theories discussed in this paper.  Concretely: while the minimal theory provides no row $z$ that satisfies the constraint $\RowPlusP {\LabTy {\Lab x} {Int}} {\LabTy {\Lab y} {Int}} z$, our Agda denotation includes both suitable instantiations for $z$ and the evidence that they satisfy the constraint.

Our claim of type soundness is semantic in nature and relies on the totality of Agda as a type theory: we show that the denotations of well-kinded types are in the denotations of their kinds, that the denotations of well-typed terms are in the denotations of their types, and so forth. Because our denotations are in a typed theory, we do not have a wrong value (as in \citet{Milner78}); instead, we extend the guarantees provided by Agda's type system to \RO.

This section gives a high-level overview of our Agda development; interested readers are referred to the full development \Supplemental.  There are two significant threads.  First: our specification of \RO so far is impredicative, while Agda is a predicative type theory.  We address this by stratifying \RO, preserving its practical expressiveness while being suitable for embedding in Agda.  Second: we need Agda definitions of the \RO primitives.  With these out of the way, the remainder of the development was pleasingly straightforward, and demonstrates soundness of kinding, typing, and equivalence.

\subsection{Stratifying \RO}

Our first challenge is developing a predicative version of \RO.  Following \citet{DunfieldK13}, we could identify the monotypes of \RO (those types without quantifiers), and limit quantifier instantiation to monotypes.  However, this approach would unacceptably compromise the expressiveness of \RO.  The following type captures a dictionary for Haskell's Monad type class:
\begin{align*}
  \TyC{Monad} &: \TypeK \to \TypeK \\
  \TyC{Monad} &= \lambda m. \, \Pi \Row {\LabTy {\Lab {return}} {\forall t \co \TypeK. \, t \to m \, t}, 
                                         \LabTy {\Lab {bind}} {\forall t, u \co \TypeK. \, m \, t \to (t \to m \, u) \to m \, u}} \\
\intertext{with selector functions such as:}                                        
  \ExC{return} &: \forall m \co \TypeK \to \TypeK, t \co \TypeK. \TyC{Monad} \, m \to t \to m \, t \\
  \ExC{return} &= \lambda d \, x. \, \ExC{sel} \, d \, \Lab {return} \, x
\end{align*}
However, to type $return$, we have to instantiate $sel$ with the type of the \Lab{return} field, $\forall t \co \TypeK, \, t \to m \, t$, which is not a monotype.

\newcommand\Level[2]{{#1}^{(#2)}}

Instead, we follow the approach of System \SF~\citep{Leivant91}, ensuring predicativity by stratifying the \RO type system.  Each type in stratified \RO is associated with a level.  We write $\Level \kappa i$ for the kinds of types at level $i$:
\begin{syntax}
  \text{Kinds} & \Level \kappa i &= \TypeK[i] \mid \LabK \mid \RowK {\Level \kappa i} \mid \Level \kappa j \to \Level \kappa k \quad \text{(where $i = j \sqcup k$)}
\end{syntax}
The base kind $\TypeK$ is now annotated with a level.  Labels are types at any level, and the types of rows and type constructors are determined by their component types.  We write $\kappa$ for the union $\bigcup_{i \in \mathbb N}{\Level \kappa i}$.

\renewcommand\KindJ[3]{#1 \vdash_{\mathrm{S}} #2 : #3}
\renewcommand\krule[1]{\labrule {k$_\textsc{s}$} {#1}}
\renewcommand\PredJ[3]{#1 \vdash_{\mathrm{S}} #2 : #3}
\newenvironment{AgdaDefs}%
  {\begin{smalle}\[\begin{array}{l@{\;}c@{\;}l} }%
  {\end{array}\]\end{smalle}\ignorespacesafterend}

\begin{figure}
\small
\begin{gather*}
\fbox{$\KindJ \Gamma \tau \kappa$}
\\
\ib{\irule[\krule{var}]
          {\EnvJ \Gamma}
          {\alpha : \kappa \in \Gamma};
          {\KindJ \Gamma \alpha \kappa}}
\rsp
\ib{\irule[\krule{$\leq$}]
          {\KindJ \Gamma \tau {\TypeK i}}
          {i \leq j};
          {\KindJ \Gamma \tau {\TypeK [j]}}}
\rsp          
\ib{\irule[\krule{$(\to)$}]
          {\EnvJ \Gamma};
          {\KindJ \Gamma {(\to)} {\TypeK [i] \to \TypeK [i] \to \TypeK [i]}}}
\\
\ib{\irule[\krule{$\I\to$}]
          {\KindJ {\Gamma, \alpha : \kappa_1} \tau \kappa_2};
          {\KindJ \Gamma {\lambda \alpha \co {\kappa_1}. \tau} {\kappa_1 \to \kappa_2}}}
\rsp
\ib{\irule[\krule{$\E\to$}]
          {\KindJ \Gamma {\tau_1} {\kappa_1 \to \kappa_2}}
          {\KindJ \Gamma {\tau_2} {\kappa_1}};
          {\KindJ \Gamma {\tau_1 \, \tau_2} {\kappa_2}}}
\rsp          
\ib{\irule[\krule{lab}]
          {\EnvJ \Gamma};
          {\KindJ \Gamma \ell \LabK}}
\\
\ib{\irule[\krule{$\then$}]
          {\PredJ \Gamma \pi i}
          {\KindJ {\Gamma, \pi} \tau \TypeK [j]};
          {\KindJ \Gamma {\pi \then \tau} \TypeK [(i + 1) \sqcup j]}}
\rsp
\ib{\irule[\krule{$\forall$}]
          {\KindJ {\Gamma, \alpha : \Level \kappa i} \tau \TypeK[j]};
          {\KindJ \Gamma {\forall \alpha\co{\Level \kappa i}. \tau} \TypeK[(i+1) \sqcup j]}}
\rsp
\ib{\irule[\krule{sing}]
          {\KindJ \Gamma \xi \LabK};
          {\KindJ \Gamma {\Sing\xi} \TypeK [0]}}
\\
\ib{\irule[\krule{lty}]
          {\KindJ \Gamma \xi \LabK}
          {\KindJ \Gamma \tau \kappa};
          {\KindJ \Gamma {\LabTy \xi \tau} \kappa}}
\rsp
\ib{\irule[\krule{row}]
          {\KindJ \Gamma \xi \LabK}
          {\KindJ \Gamma \tau \kappa};
          {\KindJ \Gamma {\Row{\LabTy \xi \tau}} {\RowK\kappa}}}
\rsp
\ib{\irule[\krule{$\Pi$}]
          {\KindJ \Gamma \rho {\RowK \kappa}};
          {\KindJ \Gamma {\Pi\rho} \kappa}}
\\
\ib{\irule[\krule{$\Sigma$}]
          {\KindJ \Gamma \rho {\RowK \kappa}};
          {\KindJ \Gamma {\Sigma\rho} \kappa}}
\isp
\ib{\irule[\krule{lift$_1$}]
          {\KindJ \Gamma \rho {\RowK{\kappa_1 \to \kappa_2}}}
          {\KindJ \Gamma \tau {\kappa_1}};
          {\KindJ \Gamma {\rho\,\tau} {\RowK{\kappa_2}}}}
\isp
\ib{\irule[\krule{lift$_2$}]
          {\KindJ \Gamma \phi {\kappa_1 \to \kappa_2}}
          {\KindJ \Gamma \rho {\RowK{\kappa_1}}};
          {\KindJ \Gamma {\phi\,\rho} {\RowK{\kappa_2}}}}
\\
\fbox{$\PredJ \Gamma \pi i$}
\\
\ib{\irule
          {\KindJ \Gamma {\rho_n} {\RowK {\Level \kappa i}}};
          {\PredJ \Gamma {\Leqp {\rho_1} {\rho_2}} i}}
\rsp
\ib{\irule
          {\KindJ \Gamma {\rho_n} {\RowK {\Level \kappa i}}};
          {\PredJ \Gamma {\RowPlusP {\rho_1} {\rho_2} {\rho_3}} i}}
\end{gather*}          
\caption{Stratified kinding and predicate formation}
\label{fig:kinding-strat}
\end{figure}

The stratified kinding relation is shown in \cref{fig:kinding-strat}.  Overall, stratification has a relatively minor impact.  Rule \krule{$\leq$} includes earlier levels in later levels; our mechanization incorporates this rule into the other rules.  Rules \krule{$\then$} and \krule{$\forall$} ensure that the result type is at least one level higher than the level of the quantified type or predicate.  The remaining rules are unchanged.  However, note that we do now require that quantification and type abstraction explicitly mention the level of the quantified or argument type.  In our mechanization, in turn, we can abstract derivations over the base level.  \Cref{fig:kinding-strat} also includes a stratified version of the predicate formation rule, tracking the level of types that appear in the predicate.


In mechanizing \RO kinds and types, we have separated the environment $\Gamma$ into three: a kinding environment $\Delta$, a predicate environment $\Phi$, and a typing environment $\Gamma$.  We use an intrinsically-kinded representation of types:

\begin{AgdaDefs}
  \mathrm{Kind} & : & \mathrm{Level} \to \mathrm{Set} \\
  \mathrm{KEnv} & : & \mathrm{Level} \to \mathrm{Set} \\
  \mathrm{Ty} & : & \forall \, \{ i \, j \co \mathrm{Level} \} \to \mathrm{KEnv} \, i \to \mathrm{Kind} \, j \to \mathrm{Set}
\end{AgdaDefs}  
We define interpretation functions for kinds, kinding environments, and types:
\begin{AgdaDefs}
  \llbracket\_\rrbracket_k & : & \forall \, \{ i \co \mathrm{Level} \} \to \mathrm{Kind} \, i \to \mathrm{Set} \, (\mathrm{lsuc} \, i) \\
  \llbracket\_\rrbracket_{ke} & : & \forall \, \{ i \co \mathrm{Level} \} \to \mathrm{KEnv} \, i \to \mathrm{Set} \, (\mathrm{lsuc} \, i) \\
  \llbracket\_\rrbracket_t & : & \forall \, \{ i \, j \co \mathrm{Level} \}  \, \{ \Delta : \mathrm{KEnv} \, i \} \{ \kappa : \mathrm{Kind} \, j \} \to \mathrm{Ty} \, \Delta \, \kappa \to \llbracket \Delta \rrbracket_{ke} \to \llbracket \kappa \rrbracket_k
\end{AgdaDefs}
These definitions are unsurprising.  For example: the kind $\TypeK [i]$ is interpreted as $Set \, i$; kinding environments are interpreted as tuples of types; the type $\forall \alpha \co \kappa. \tau$ in kinding environment $H$ is interpreted as a dependent function $(X : \llbracket \kappa \rrbracket_k) \to \llbracket \tau \rrbracket_t (H, X)$.  Label singleton types are all interpreted as $\top$ (the unit type), buttressing our claim that \RO can be implemented without runtime manipulation or comparison of labels.  The interpretation of row types, records, and variants is discussed next.

The interpretation of types gives a constructive proof of the following claim:
\begin{theorem*}
  The kind system of \RO is sound.
\end{theorem*}

\noindent
Of course, this is only convincing if the interpretations themselves are non-trivial.  Here we rely on the underlying type theory: for example, as we interpret the kind of type constructors $\kappa_1 \to \kappa_2$ as Agda functions $\llbracket \kappa_1 \rrbracket_k \to \llbracket \kappa_2 \rrbracket_k$, we can be confident that our interpretations of types of that kind are meaningful.  For the full details, please see the Agda development \Supplemental.

\subsection{Rows and Indices}

We intend our interpretation of records and variants to be both type-safe, and to align with the intuition of those types.  That is, a record should be a sequence of its field values, and a variant should be a single tagged value.  

We begin with rows themselves.  Intuitively, a row is a sequence of types.  Our encoding in Agda is almost that direct:
\begin{AgdaDefs}
  \mathrm{Row} & : & \forall \{ i : \mathrm{Level} \} \to \mathrm{Set} \, i \to \mathrm{Set i} \\
  \mathrm{Row} \, A & = & \Sigma[n \in \mathbb N] (\mathrm{Fin} \, n \to A)
\end{AgdaDefs}
That is to say: a row at level $i$ is a dependent pair of its length $n$ and a map from finite indices less than $n$ to types at level $i$.  We can define record and variant constructors (at type $\TypeK [i]$) as dependent functions on rows:
\begin{AgdaDefs}
  \Pi & : & \forall \{ i : \mathrm{Level} \} \to \mathrm{Row} \, (\mathrm{Set} \, i) \to \mathrm{Set} \, i \\
  \Pi \, (n, P) & = & (i : \mathrm{Fin} \, n) \to P \, i \\[1ex]
  \Sigma & : & \forall \{ i : \mathrm{Level} \} \to \mathrm{Row} \, (\mathrm{Set} \, i) \to \mathrm{Set} \, i \\
  \Sigma \, (n, P) & = & i : \Sigma[ i \in \mathrm{Fin} \, n] (P \, i)
\end{AgdaDefs}

(We will rely on some overloading to avoid tedious qualified names: $\Sigma$ followed by a variable binding is the dependent sum constructor; followed by a row, it is the variant constructor.)  In each case, we pattern match on the input row, obtaining its length $n$ and a mapping from indices to types $P$.  A variant is the expected tagged value, pairing a tag less than n with a value of the type indexed by the tag.  A record is another dependent function: given an index into the record, it returns a value of the type at that index.

We have made one simplification relative to \RO: we implement records and variants only at the base kind, and express the type constructor variants using type functions.  This does not reflect a fundamental limitation in our embedding, but simply a choice made for expediency in development.

These definitions emulate our intuition of records and variants.  For variants, we are quite close: erasing the types leaves a pair of a tag and a value, just as you might expect to represent a value of a traditional variant type.  For records, we are further away: while we emulate accessing fields of a record by offset, the practical construction of records is not emulated by our encoding.  Nevertheless, we hope that these encodings demonstrate the potential of a real implementation, even if they do not claim to address all the problems that such an implementation would encounter.

Note that these types have none of the properties we have assumed for the corresponding types in \RO: there are no traces of labels to be found, and order is very much significant in determining the meaning of rows, records, and variants.  The mapping between rows in the source language and rows in Agda will be found in the concrete evidence for the row predicates, discussed next.

\subsection{Containment and Combination}

The next piece of our encoding is the evidence for the containment and combination predicates.  The stratification of the entailment relation $\EntJS \Mty \Gamma \pi$ is entirely unsurprising.  As usual in qualified types, evidence for predicates plays a central role in interpreting the overloaded operators.  Pur goal is to combine the intuition of a practical realization of \RO with dependent types to ensure type safety.

Intuitively, containment maps indices in the smaller row to indices in the larger row.
\begin{AgdaDefs}
  \_{\lesssim}\_ & : & \forall \{ i \co \mathrm{Level} \} \, \{ A \co \mathrm{Set} \, i \} \to \mathrm{Row} \, A \to \mathrm{Row} \, A \to \mathrm{Set} \, i \\
  (n, P) \lesssim (m, Q) & = & (i \in \mathrm{Fin} \, n) \to \Sigma[j \in \mathrm{Fin} \, m] (P \, i \equiv Q \, j)
\end{AgdaDefs}
(We have omitted some straightforward but tedious bookkeeping to do with levels.)  The evidence for containment is a dependent function over indices in the smaller row, associating each with both an index in the larger row and a proof that the associated types are the same.  Implementing record projection and variant injection in terms of this evidence is simple: the former simply precomposes with the evidence function while the latter replaces the existing tag with its image in the evidence function.

Similarly, combination maps indices in the resulting row to indices in one of the two starting rows:
\begin{AgdaDefs}
  \_{\odot}\_{\sim}\_ & : & \forall \{ i \co \mathrm{Level} \} \, \{ A \co \mathrm{Set} \, i \} \to \mathrm{Row} \, A \to \mathrm{Row} \, A \to \mathrm{Row} \, A \to \mathrm{Set} \, i \\
  (l, P) \odot (m, Q) \sim (n, R) & = & (i \in \mathrm{Fin} \, n) \to (\Sigma[ j \in \mathrm{Fin} \, l] (P \, j \equiv R \, i)) \mathrel{\mathrm{or}} (\Sigma[ j \in \mathrm{Fin} \, m] (Q \, j \equiv R \, i))
\end{AgdaDefs}

We pair the intuitive mapping on indices with evidence that types agree.  As for containment, the implementation of the branching and concatenation operators in terms of this evidence is immediate.  Unfortunately, however, this is not sufficient to implement all of the entailment rules of \RO.  Our intuition is not just that this be \emph{any} map between the indices, but a \emph{surjective} map: every index in one of the original rows should appear somewhere in the combined row.  This intuition justifies the entailment rules \entrule{${\odot}{\lesssim}$l} and \entrule{${\odot}{\lesssim}$r}, which conclude containment from combination.  However, this intuition is not captured in our evidence.  We have taken a brute force approach to doing so, by storing the evidence for the two containments in the evidence for combination:
\begin{AgdaDefs}
  (l, P) \odot (m, Q) \sim (n, R) & = & (i \in \mathrm{Fin} \, n) \to (\Sigma[ j \in \mathrm{Fin} \, l] (P \, j \equiv R \, i)) \mathrel{\mathrm{or}} (\Sigma[ j \in \mathrm{Fin} \, m] (Q \, j \equiv R \, i)) \\
  & & {} \times (l, P) \lesssim (n, R) \times (m, Q) \lesssim (n, R)
\end{AgdaDefs}
This definition allows us to realize all of \RO's entailment rules.

We define an intrinsically-kinded representation of predicates, interpreted as evidence:
\begin{AgdaDefs}
  \mathrm{Pred} & : & \forall \, \{ i \, j \co \mathrm{Level} \} \to \mathrm{KEnv} \, i \to \mathrm{Kind} \, j \to Set \\
  \llbracket\_\rrbracket_p & : & \forall \{ i \, j \co \mathrm{Level} \} \, \{ \Delta : \mathrm{KEnv} \, i \} \, \{ \kappa : \mathrm{Kind} \, j \} \to \mathrm{Pred} \, \Delta \, \kappa \to \llbracket \Delta \rrbracket_{ke} \to \mathrm{Set} \, (\mathrm{lsuc} \, j)
\end{AgdaDefs}
We define a corresponding intrinsically well-formed definition of predicate environments and entailment:
\begin{AgdaDefs}
  \mathrm{PEnv} & : & \forall \{ i \co \mathrm{Level} \} \to \mathrm{KEnv} \, i \to \mathrm{Level} \to \mathrm{Set} \\
  \mathrm{Ent} & : & \forall \, \{ i_1 \, i_2 \, i_3 \co \mathrm{Level} \} \{ \kappa \co \mathrm{Kind} \, i_3 \} \to
                     (\Delta \co \mathrm{KEnv} \, i_1) \to \mathrm{PEnv} \, \Delta \, i_2 \to \mathrm{Pred} \, \Delta \, \kappa \to \mathrm{Set}
\end{AgdaDefs}
Finally, we define the meaning of an entailment judgment in terms of the meaning of the predicate it entails:
\begin{AgdaDefs}
  \llbracket\_\rrbracket_{pe} & : & \forall \, \{ i \, j \co \mathrm{Level} \} \{ \Delta \co \mathrm{KEnv} \, i \} \to
                                    \mathrm{PEnv} \, \Delta \, j \to \llbracket \Delta \rrbracket_{ke} \to \mathrm{Set} \, j \\
  \llbracket\_\rrbracket_{n} & : & \forall \, \{ i_1 \, i_2 \, i_3 \co \mathrm{Level} \} \{ \kappa \co \mathrm{Kind} \, i_3 \} \,
                                              \{ \Delta \co \mathrm{KEnv} \, i_1 \} \, \{ \Phi \co \mathrm{PEnv} \, \Delta \, i_2 \} \,
                                              \{ \pi \co \mathrm{Pred} \, \Delta \, i_3 \} \to \\
                             & &   \quad \mathrm{Ent} \, \Delta \, \Phi \, \pi \to (H : \llbracket \Delta \rrbracket_{ke}) \to
                                     \llbracket \Phi \rrbracket_{pe} \, H \to \llbracket \pi \rrbracket_{p} \, H
\end{AgdaDefs}
The latter provides a constructive proof of the following.
\begin{theorem*}
  The entailment relation of \RO is sound.
\end{theorem*}

\subsection{Label-Generic Operations}

The label-generic operators $\Ana$, $\Syn$, and $\Fold$ work by invoking a suitably parametric function on entries in their source rows.  To implement this, we must be able to work backwards from the index used in a variant or record to the corresponding evidence that its type is in the original row.  We capture this in Agda as follows.

We begin by introducing an abbreviation for indices over a given row.
\begin{AgdaDefs}
  \mathrm{Ix} & : & \forall \, \{ i : \mathrm{Level} \} \, \{ A : \mathrm{Set} \, i \} \to \mathrm{Row} \, A \to \mathrm{Set} \\
  \mathrm{Ix} \, (n, \_) & = & \mathrm{Fin} \, n 
\end{AgdaDefs}
The $pick$ operator selects from a row the singleton row at a particular index, and we can construct evidence that each singleton row is contained within the original row.
\begin{AgdaDefs}
  \_\mathrm{pick}\_ & : & \forall \, \{ i : \mathrm{Level} \} \, \{ A : \mathrm{Set} \, i \} \to (\rho \co \mathrm{Row} \, A) \to \mathrm{Ix} \, \rho \to \mathrm{Row} \, A \\
  \mathrm{pickedIn} & : & \forall \, \{ i : \mathrm{Level} \} \, \{ A : \mathrm{Set} \, i \} \, \{\rho \co \mathrm{Row} \, A \} \, \{ n : \mathrm{Ix} \, \rho \} \to \rho \mathbin{\mathrm{pick}} n \lesssim \rho
\end{AgdaDefs}
Similarly, the $delete$ operator returns the row containing everything but the given index.  We can also construct evidence that this row is contained within the original.
\begin{AgdaDefs}
  \_\mathrm{delete}\_ & : & \forall \, \{ i : \mathrm{Level} \} \, \{ A : \mathrm{Set} \, i \} \to (\rho \co \mathrm{Row} \, A) \to \mathrm{Ix} \, \rho \to \mathrm{Row} \, A \\
  \mathrm{deletedIn} & : & \forall \, \{ i : \mathrm{Level} \} \, \{ A : \mathrm{Set} \, i \} \, \{\rho \co \mathrm{Row} \, A \} \, \{ n : \mathrm{Ix} \, \rho \} \to \rho \mathbin{\mathrm{delete}} n \lesssim \rho
\end{AgdaDefs}
Finally, for a given index into a row, we can produce the evidence needed to invoke the body of a label-generic operator: that combining the singleton row and that index and the remainder of the row gives the original row.
\begin{AgdaDefs}
  \mathrm{recombine} & : & \forall \, \{ i : \mathrm{Level} \} \, \{ A : \mathrm{Set} \, i \} \to (\rho \co \mathrm{Row} \, A) \to (n : \mathrm{Ix} \, \rho) \to \rho \mathbin{\mathrm{pick}} n \odot \rho \mathbin{\mathrm{delete}} n \sim \rho
\end{AgdaDefs}
The implementations of the label-generic operators follow easily.

\subsection{Terms and Equivalences}

Finally, we come to the representations of terms, and of type equivalences.  We use intrinsically kinded representations of type and predicate equivalence:
\begin{AgdaDefs}
  \_{\equiv_p}\_ & : & \forall \, \{ \Delta \co \mathrm{KEnv} \} \, \{ \kappa \co \mathrm{Kind} \} \to \mathrm{Pred} \, \Delta \, \kappa \to \mathrm{Pred} \, \Delta \, \kappa \to \mathrm{Set} \\
  \_{\equiv_t}\_ & : & \forall \, \{ \Delta \co \mathrm{KEnv} \} \, \{ \kappa \co \mathrm{Kind} \} \to \mathrm{Ty} \, \Delta \, \kappa \to \mathrm{Ty} \, \Delta \, \kappa \to \mathrm{Set} \\
\end{AgdaDefs}
(We will omit the level bookkeeping for the remainder of this section, as it is entirely routine.)

We have made one important simplification in mechanizing the type equivalence relation.  If we restrict type equivalence to kinds $\TypeK [i]$, then we have shown that the interpretation of equivalence derivations is an isomorphism in Agda.  That is to say, if we have a derivation that $\tau_1 \equiv \tau_2$, then (for a suitable type environment $H$) we can show not only functions $to : \llbracket \tau_1 \rrbracket H \to \llbracket \tau_2 \rrbracket H$ and $from : \llbracket \tau_2 \rrbracket H \to \llbracket \tau_1 \rrbracket H$, but also that their compositions are the identity function.  In particular, we validate rule \erule{sing}, that singleton record and variant types are isomorphic to their underlying field type.

However, this definition of isomorphism is not applicable at higher kinds: type constructors have no elements, so it makes little sense to talk about mappings between them.  Moreover, if we remove rule \erule{sing}, we are able to show stronger results, which generalize to all kinds:
\begin{AgdaDefs}
  \llbracket\_\rrbracket_{ep} & : & \forall \{ \Delta \co \mathrm{KEnv} \} \, \{ \kappa \co \mathrm{Kind} \} \, \{ \pi_1 \, \pi_2 \co \mathrm{Pred} \, \Delta \, \kappa \} \to \pi_1 \equiv_p \pi_2 \to (H : \llbracket \Delta \rrbracket_{ke}) \to \llbracket \pi_1 \rrbracket_p \, H \equiv \llbracket \pi_2 \rrbracket_p \, H \\
  \llbracket\_\rrbracket_{et} & : & \forall \{ \Delta \co \mathrm{KEnv} \} \, \{ \kappa \co \mathrm{Kind} \} \, \{ \tau_1 \, \tau_2 \co \mathrm{Ty} \, \Delta \, \kappa \} \to \tau_1 \equiv_t \tau_2 \to  (H : \llbracket \Delta \rrbracket_{ke}) \to \llbracket \tau_1 \rrbracket_t \, H \equiv \llbracket \tau_2 \rrbracket_t \, H
\end{AgdaDefs}
That is to say: we show that when equivalence is derivable between two predicates or two types (at any kind), their interpretations are propositionally equal in Agda.  Given our limitations, these provide constructive proofs of the following claim.
\begin{theorem*}
  The type and predicate equivalence relations of \RO are sound.
\end{theorem*}

To account for the loss of \erule{sing}, our term language is extended with terms to construct and deconstruct singleton records and variants.  We define intrinsically-typed representations of terms, and their interpretation:
\begin{AgdaDefs}
  \mathrm{Env} & : & \mathrm{KEnv} \to \mathrm{Set} \\
  \mathrm{Tm} & : & (\Delta \co \mathrm{KEnv}) \to \mathrm{PEnv} \, \Delta \to \mathrm{Env} \, \Delta \to \mathrm{Ty} \, \Delta \, \TypeK \to \mathrm{Set} \\[1ex]
  \llbracket\_\rrbracket_{e} & : & \forall \{ \Delta \co \mathrm{KEnv} \} \to \mathrm{Env} \, \Delta \to \llbracket \Delta \rrbracket_{ke} \mathrm{Set} \\
  \llbracket\_\rrbracket_{t} & : & \forall \{ \Delta \co \mathrm{KEnv} \} \, \{ \Phi \co \mathrm{PEnv} \} \, \{ \Gamma \co \mathrm{Env} \} \, \{ \tau : \mathrm{Ty} \,  \Delta \, \TypeK \} \to \\
    & & \quad \mathrm{Tm} \, \Delta \, \Phi \, \Gamma \, \tau \to (H \co \llbracket \Delta \rrbracket_{ke}) \to \llbracket \Phi \rrbracket_{pe} \, H \to \llbracket \Gamma \rrbracket_{e} \, H \to \llbracket \tau \rrbracket_{t} \, H
\end{AgdaDefs}
The latter provides a constructive proof of our final claim.
\begin{theorem*}
  The type system of \RO is sound.
\end{theorem*}

\section{Related Work}
\label{sec:related}

There is a significant and growing literature on row types and their applications, and a larger literature on extensible data types in general.  We highlight that work that is most relevant to \RO.

\subsubsection*{Featherweight Ur.}
\newcommand\citeposs[1]{\citeauthor{#1}'s~\citeyear{#1}}

The most immediately relevant languages are Featherweight Ur and its practical realization in Ur/Web~\citep{Chlipala10,Chlipala15a,Chlipala15b}.  As in \Rose, Ur supports row and record concatenation with first-class labels, enabled by first-class label inequality proofs.  As for \RO, Ur is based on System~\FO, and supports mapping type-level operations over rows.  Ur has practical evaluation as a framework for database-connected web applications.  We view \RO and Ur as complementary explorations of the design space of extensible data types.

There are several differences in focus between \RO and Ur.  Ur does not include extensible variants.  Consequently, the duality of records and variants does not appear in Ur, and examples like our $reify$ and $reflect$ functions do not apply to Ur.  We view extensible variants as an important application of row typing, useful for examples like the expression problem and encoding extensible effects; however, we do not think there is any fundamental reason that Ur's approach to extensible records could not be equally applicable to extensible variants.  Ur also does not attempt to generalize over different row theories, but assumes that row disjointness is sufficient to capture extensibility.

The more significant difference between \RO and Ur is in our approach to generic programming with records.  Ur provides a family of folding functions for concrete records types.  Instead of our view, in which folds should respect the identities of the underlying row theory, Ur uses the type of its folder to capture the particular order in which the programmer intends to visit fields in the records.  We believe that \RO's \emph{syn}thesis operator provides a novel, alternative view of generic programming with records.  In particular, we are able to define many of our operations to apply to records regardless of their structure; while we believe that Ur's folder could capture the same operation for any concrete record type, it is less clear that Ur captures them in the general case.

\subsubsection*{Other row type systems.}

Row types were originally proposed by \citet{Wand87} as a mechanism for typing records and variants; he defined rows by extension, one field at a time, and allowed subsequent extensions to overwrite fields already in a record.  \citet{Remy89} generalized Wand's approach in several significant ways.  He restricts row extension to fields not already present in the row, enforced using kinds.  His rows record both present and absent fields, with explicit operations to ``forget'' entries in rows.  Finally, he introduces polymorphism over field presence, allowing his calculus to capture patterns like a single operation for both record extension and record update.  \Remy's approach has been used as the foundation for numerous other row type systems.  \citet{BlumeAC06} extends \Remy's approach to incorporate first-class blocks over extensible variants.  Their implementation relies on the duality between records and variants, translating case blocks into records.  However, this duality is not exposed to the programmer; unlike \RO, they rely on having a specific type for case blocks distinct from the normal function type.  Other application of \Remy-style row type systems include: \citeposs{MakholmW05} system for mixin modules; \citeposs{LindleyC12} type system for effect polymorphism; \citeposs{HillerstromL16} system for extensible effects and handlers; and, \citeposs{LindleyM17} account of extensible session types.  \citet{GasterJ96} implement a system with operations similar to \Remy's, but using qualified types instead of kinds to assure that row extension is well-defined.  \citet{LindleyMM17} start from \Remy-style rows, but consider several extensions including generic support for renaming entries in rows.  \citet{BerthomieuM95} and \citet{Leijen05} independently proposed scoped rows, in which row extension preserves both the original and new fields.

\citet{Wand91} identified the problems that can arise in typing record concatenation, and proposed an approach based on intersection types.  \citet{HarperP91} support record concatenation using a new form of quantification, in which quantification is over types disjoint from a given row.  Their system cannot express Wand's problem: while it can require that two rows be disjoint, it cannot require that a single field appear in their concatenation without requiring that it appear in a particular input row.

There have been numerous encodings of row types in other type system features, most notably Haskell's type classes and type families~\citep{KiselyovLS04,Swierstra08,Bahr14,Morris15,OliveiraMY15}.  While impressive, these encodings inevitably rely on encoding rows as particular sequences as types, and so struggle to capture the flexibility that row typing is intended to provide.  Extensible data types can also be expressed directly using intersection types and the merge operator~\cite{Dunfield12,RiouxHOZ23}.  

\RO is differentiated from other row type theories by its focus on label-generic operations.  It also inherits the expressiveness of \Rose, and its adaptability to multiple different row theories.

\subsubsection*{Shallow embeddings in Agda.}

Our approach to mechanizing the metatheory of \RO is unusual; far more typical would have been to define an operational semantics of \RO directly, and then mechanize the expected properties of that operational semantics.  We chose to embed the semantics of \RO directly in Agda for two reasons: we wanted an account that clearly did not rely on labels themselves, and we needed to rely on dependent typing to guarantee that record and variant operations were well-typed.  This made it natural to embed our semantics in a dependent type theory, and Agda provides flexible dependently-typed programming and a rich standard library.

Embedding simply-typed $\lambda$-calculi in rich type theories is well-traveled ground.  There is recent work on shallow or mixed deep and shallow embeddings of rich type theories in rich type theories~\cite{McBride10,KaposiKK19}.  Our embedding is less impressive than theirs: while we demonstrate that our notions of type equality and predicate entailment are sound, we still require explicit equality and entailment proofs in our derivations.  

\section{Conclusion}
\label{sec:conclusion}

We have presented a novel approach to programming with extensible data types, based on label-generic operators for variant destruction and record construction and destruction.  We conclude by identifying several directions of future work.

\paragraph{Relating row components.}

\citet{LindleyMM17} proposes a renaming operator for rows, as a tool for simulating scoped rows with simple rows.  We might hope to capture such an idea in \RO; indeed, our kind system even includes rows of labels, which seem like a promising start.  However, while we could attempt to describe a function that relabeled the fields of a row or constructors of a variant, we have no way to guarantee that the renamed fields are unique!  That is, we have nothing that accepts the row $\Row{\LabTy{\Lab a} {\Lab b}, \LabTy {\Lab b} {\Lab c}}$ while rejecting the row $\Row{\LabTy {\Lab a} {\Lab z}, \LabTy {\Lab b} {\Lab z}}$.  More generally, we have no way to impose conditions on the relationship between an entry in a row and the remainder, other than that provided by the row combination predicate.

\paragraph{Realizing \RO.}

\RO's goal is to demonstrate the expressiveness of its core features.  We identify two challenges in making \RO more practical.  The first is exposing its features in a programmer-friendly surface language, such as a variant of Haskell.  Doing so would allow us to use \RO to capture practical examples from algebraic effects and handlers to extensible compiler passes.  While adapting \RO to a type system without type-level functions would certainly make type reconstruction more likely, it may also introduce limitations in \RO's expressiveness.  The second is an efficient implementation of extensible records and variants, in particular, an account of record construction that does not require copying record values or leave records fragmented.

\begin{acks}
  We thank: James McKinna, for providing initial direction to our mechanization of \RO as well as general discussion of \Rose; Christa Jenkins, for guidance in developing the mechanization; and, Fabian Ruch for extensive feedback on the final mechanization.   This work was supported by the \grantsponsor{}{National Science Foundation}{https://www.nsf.org} under Grant No.~\grantnum{}{CCF-2044815}.
\end{acks}

\section*{Data Availability Statement}
\label{sec:data}
Our Agda mechanization of \RO is available online \citep{artifact}.

\bibliographystyle{ACM-Reference-Format}
\bibliography{strings,higher}

\ifextended

\clearpage

\appendix

\section{Typing Derivations of Sample \RO terms}
\label{sec:derivations}

\renewcommand\KindJ[3]{#1 \vdash #2 : #3}
\renewcommand\PredJ[2]{#1 \vdash #2}
\begin{figure}[H]
{\small
\begin{gather*}
\ib{
  \irule
    {\irule
      {\irule
        {\irule
         {\irule {...} ; {\PredJ {\Gamma_2} {\Leqp {\Row {\LabTy l t}} {z}}}}
         {\irule
           {\irule
             {{\irule
                 {\irule
                   {\TypeJ {\Gamma_5} {r} {\Pi z}} {} ; {\TypeJ {\Gamma_5} {{\Prj r}} {\Row {\LabTy l t}}}}
                 {\irule {} ; {\TypeJ {\Gamma_5} {g} {\Sing l}}} ;
                 {\TypeJ {\Gamma_4, g \co \Sing l} {\Unlabel {\Prj r} g} {t}}}}
             ; {\TypeJ {\Gamma_3, r \co \Pi z} {\lambda (g : \Sing l). \, \Unlabel {\Prj r} g} {\Sing l \to t}}} 
            ; {\TypeJ {\Gamma_2, \Leqp {\Row {\LabTy l t}} {z}}
              {\lambda (r \co \Pi z) \, (g : \Sing l). \, \Unlabel {\Prj r} g}
              {\Pi z \to \Sing l \to t}}} ; 
        {\TypeJ {\Gamma_1, z \co \RowK \TypeK}
    {\, \lambda (r \co \Pi z) \, (g : \Sing l). \, \Unlabel {\Prj r} g}
    {\Leqp {\Row {\LabTy l t}} z \then \Pi z \to \Sing l \to t}}} ;
      {\TypeJ {\Gamma_0, t \co \TypeK}
    {\Lambda (z \co \RowK \TypeK). \, \lambda (r \co \Pi z) \, (g : \Sing l). \, \Unlabel {\Prj r} g}
    {\forall z \co \RowK \TypeK. \, \Leqp {\Row {\LabTy l t}} z \then \Pi z \to \Sing l \to t}}} ;
    {\TypeJ {\varepsilon , l \co \LabK}
    {\Lambda (t \co \TypeK) \, (z \co \RowK \TypeK). \, \lambda (r \co \Pi z) \, (g : \Sing l). \, \Unlabel {\Prj r} g}
    {\forall t \co \TypeK, z \co \RowK \TypeK. \, \Leqp {\Row {\LabTy l t}} z \then \Pi z \to \Sing l \to t}}} ;
  {\TypeJ \varepsilon
    {\Lambda (l \co \LabK) \, (t \co \TypeK) \, (z \co \RowK \TypeK). \, \lambda (r \co \Pi z) \, (g : \Sing l). \, \Unlabel {\Prj r} g}
    {\forall l \co \LabK, t \co \TypeK, z \co \RowK \TypeK. \, \Leqp {\Row {\LabTy l t}} z \then \Pi z \to \Sing l \to t}}}
\end{gather*}}
\end{figure}
\fi

\end{document}